\documentclass[fleqn,usenatbib]{mnras}

\usepackage{newtxtext,newtxmath}

\usepackage[T1]{fontenc}

\DeclareRobustCommand{\VAN}[3]{#2}
\let\VANthebibliography\thebibliography
\def\thebibliography{\DeclareRobustCommand{\VAN}[3]{##3}\VANthebibliography}

\usepackage{graphicx}	
\usepackage{amsmath}	
\usepackage{float}      
\usepackage{tabularx}

\title[IC 2574 Globular Cluster Candidates]{Discovery of Globular Cluster Candidates in the Dwarf Irregular Galaxy IC 2574 Using HST/ACS Imaging}

\author[N. Karim et al.]{Noushin Karim$^{1}$,
Michelle L. M. Collins$^{1}$, Duncan A. Forbes$^{2}$ and Justin I. Read$^{1}$\\
$^{1}$Physics Department, University of Surrey, Guildford, GU2 7XH, UK\\
$^{2}$Centre for Astrophysics and Supercomputing, Swinburne University, Hawthorn, VIC 3122, Australia\\}

\pubyear{2023}

\begin{document}
\label{firstpage}
\pagerange{\pageref{firstpage}--\pageref{lastpage}}
\maketitle
\begin{abstract}
We report the discovery of 23 globular cluster (GC) candidates around the relatively isolated dwarf galaxy IC 2574 within the Messier 81 (M81) group, at a distance of 3.86 Mpc. We use observations from the HST Advanced Camera for Surveys (ACS) to analyse the imaging in the F814W and F555W broadband filters. Our GC candidates have luminosities ranging from $-5.9 \geq M_V \geq -10.4$ and half-light radii of $1.4 \leq r_h \leq 11.5$~pc. We find the total number of GCs ($N_{\mathrm{GC}})=27\pm5$ after applying completeness corrections, which implies a specific frequency of $S_N = 4.0\pm0.8$, consistent with expectations based on its luminosity. The GC system appears to have a bimodal colour distribution, with 30\% of the GC candidates having redder colours. We also find 5 objects with extremely blue colours that could be young star clusters linked to an intense star formation episode that occurred in IC 2574 $\sim$1 Gyr ago.
We make an independent measurement of the halo mass of IC 2574 from its kinematic data, which is rare for low mass galaxies, and find log $M_{200} = 10.93 \pm 0.08$. We place the galaxy on the well-known GC system mass-halo mass relation and find that it agrees well with the observed near-linear relation. IC 2574 has a rich GC population for a dwarf galaxy, which includes an unusually bright $\omega$ Cen-like GC, making it an exciting nearby laboratory for probing the peculiar efficiency of forming massive GCs in dwarf galaxies.
\end{abstract}


\section{Introduction}
\label{sec:intro}

Globular clusters (GCs) are among the oldest astronomical objects to exist, providing a unique probe into the early Universe and the conditions under which the first galaxies formed, making them reliable tracers of galaxy formation and evolution \citep{Puzia_2005,Strader_2005,Kruijssen_2014,Orkney_2019,Beasley_2020}. GCs are composed of a homogeneous stellar population with similar ages, metallicities and chemical compositions, although they do show interesting spreads and correlations in their light element abundances \citep[][and references therein]{Osborn_EA_1971,Shetrone_EA_1996,Ivans_EA_1999,Yong_EA_2003,Gratton_2004,Piotto_2007,D_Antona_2007,Carretta_EA_2009,Marino_EA_2012,Meszaros_EA_2015,Bastian_2018,Nataf_EA_2019}.
Despite complex galaxy evolution through mergers and interactions, high density GC cores can remain largely intact and retain unique characteristics of the environment in which they were born \citep[][and references therein]{Penarrubia_2009,Mucciarelli_2021}. This makes them a promising way to gauge the evolution of galaxies over their lifetimes \citep{Kissler-Patig_2000}. 
The properties of GC systems as a whole have a remarkable story to tell in this context. The number of GCs ($N_{\mathrm{GC}}$) in a system allow us to make an estimate of the GC system mass ($M_{\mathrm{GC}}$) \citep[e.g.][]{McLaughlin_2005,Spitler_2009}, which is found to follow tight scaling relations with the halo mass and total luminosity of the host galaxy. This physical link between the properties of the GC system and the host galaxy was first established by \citet{Blakeslee_1997}, and has grown in to a rich topic of study using various host galaxy properties \citep{Santiago_1993,Zepf_1993,Peng_2008,Spitler_2009,Georgiev_2010,Harris_2013,Harris_2015,Hudson_2014,Forbes_2018_Extending,Forbes_Remus_2018,Burkert_Forbes_2020,Beasley_2020,Eadie_2022,Dornan_2023}. 
These scaling relations suggest that GCs formed in the smallest proto-galactic building blocks \citep[e.g.][]{Boley_2009} and therefore accurately track the assembly history of galaxies thereafter \citep{Harris_1991,Ashman_Zepf_1998_BOOK,El_Badry_2019,Beasley_2020}. $N_{\mathrm{GC}}$, $M_{\mathrm{GC}}$, the luminosity, size and colour distribution all offer insightful clues about the history of their host galaxies \citep[][and references therein]{Kissler-Patig_2000, Spitler_2009,Georgiev_2010,Hudson_2014,Harris_2017,Amorisco_2018,Forbes_2018_Extending,Prole_2019,Beasley_2020,Doppel_2021,TrujilloGomez_2021}. 

$M_{\mathrm{GC}}$ is found to have an empirical, near-linear correlation with $M_{200}$ across 7 orders of magnitude \citep{Spitler_2009,Georgiev_2010,Hudson_2014,Harris_2017,Forbes_2018_Extending}.
Populating the $M_{\mathrm{GC}} - M_{200}$ relation with observed galaxy data is vital to build a complete picture, however measuring the halo mass of individual galaxies is difficult and relies on kinematic data from the galaxy. A review by \citet{Courteau_2014} discusses the variety and reliability of mass estimator techniques in detail. However, each of these have their own limitations \citep{Courteau_2014,Coles_2014,Read_2017}, and often require expensive and targeted follow-up observations to determine precision masses. The $M_{\mathrm{GC}} - M_{200}$ relation implies that by finding the GC system mass of a galaxy, one can statistically infer the galaxy's halo mass, circumventing the difficulties associated with directly observing and measuring the halo mass; although for $N_{\mathrm{GC}} < 15$ there is considerable scatter in the relation, with little predictive power \citep{Burkert_Forbes_2020}.\\

In this paper, we investigate the GC system of IC~2574, a particularly intriguing, gas-rich, nearby and relatively isolated dwarf irregular (dIrr) galaxy in the M81 system.
IC~2574 has a large effective radius, $r_e$ = 2.7 kpc, and a low surface brightness, $\mu_{r',0}$ = 22.6~mag~arcsec$^{-2}$, which is similar to, but more luminous than, an ultra diffuse galaxy (UDG); UDGs are known to have significant variations in their GC populations \citep{Forbes_UDGs_2020}.

IC~2574 is at a distance of 3.86 Mpc \citep{Dalcanton_2009} and is located in the outer regions of the M81 system; it is believed to have not undergone any recent interactions, making it unlikely to have been influenced by external tidal forces \citep{Pellerin_2012,Sorgho_2020,Mondal_2021}. IC~2574 has experienced a constant star formation history (SFH) with an intense star-forming event that occurred 1~Gyr ago \citep{Weisz_2008}. IC~2574's circular velocity ($v_{\mathrm{max}}$) from \citet{Oh_2011} allows us to calculate an independent measurement of its halo mass, which is rare for low mass galaxies \citep{Forbes_2018_Extending}. Thus, a search for the GC candidates of IC~2574 allows us to place another galaxy on the $M_{\mathrm{GC}} - M_{200}$ relation and determine whether IC~2574 agrees with the observed near-linear relation \citep{Spitler_2009}.

IC~2574 appears to host a rich GC population and based on its stellar mass it should, although this has never been assessed in detail. \citet{Georgiev_1996} presented results of a GC candidate search around IC~2574, identifying 11 candidates using photographic plates, including a cluster as luminous as the unusual Milky Way (MW) GC $\omega$ Cen. It is the most luminous GC in the MW \citep{Mackey_2005} and is thought to be the remnant of a tidally stripped dwarf galaxy \citep{Majewski_2000,Hilker_2000,Leaman_2012}.
It would be unexpected to find an analogue of $\omega$ Cen in IC~2574 as there is no evidence for recent mergers in this galaxy \citep{Pellerin_2012,Mondal_2021}; as such, this could give us insight into a possible major merger in IC~2574's early history, or alternatively help us to understand what conditions aid the peculiar efficiency of forming massive GCs in dwarf galaxies \citep[e.g.][]{Larsen_NGC4214_NGC4449_2004,Sharina_2005,Villegas_2008,deGrijs_NGC5253_2013}.
In the analysis by \citet{Georgiev_1996}, magnitudes and colours were presented for the GC candidates, but not coordinates.\\ 

For the first time since, we perform a GC candidate search around IC~2574 using deep HST/ACS imaging to establish its GC system, and validate the claims of an $\omega$ Cen-like candidate.

The outline of this paper is as follows: in Section \ref{sec:data&method} we describe where we obtained the data and outline the method of photometry, the GC candidate selection criteria and the completeness corrections, in Section \ref{sec:results} we present our candidate GC system and the resulting GC system-galaxy scaling relations, and finally in Section \ref{sec:conclusion} we summarise our findings and discuss future work.

\begin{table}
	\centering
	\caption{Properties of IC~2574: (a)~\citet{Wenger_SIMBAD_2000}, (b)~\citet{Dalcanton_2009}; distance calculated and uncertainty propagated from distance modulus $27.93 \pm 0.03$, (c)~\citet{Chiboucas_2009}, (d)~\citet{Wenger_SIMBAD_2000,Cook_2014}, (e)~\citet{Cannon_2005}, (f)~\citet{Pasquali_2008}, (g)~\citet{Lee_2006}, (h)~\citet{SFD_1998}, (i)~\citet{Oh_2011}; maximum uncertainty assumed from velocity difference (true and derived rotation velocities) due to instrument limitations and observational uncertainties. Values without references are calculated in this work. Uncertainty in $M_V$ is propagated from uncertainty in the distance modulus and uncertainty in $V$ \citep{Wenger_SIMBAD_2000}. Uncertainty in log$M_{200}$ is propagated from the uncertainty in $v_{\mathrm{max}}$. Uncertainty in $N_{\mathrm{GC}}$ is Poissonian.}
	\label{tab:ic2574properties}
	\begin{tabular}{ll}
             \\
		\hline
            \hline
		Properties & Value\\
		\hline
            \hline
		RA (J2000) & 10:28:23.61$^{\hspace{0.03cm} a}$\\
            Dec (J2000) & +68:24:43.44$^{\hspace{0.03cm} a}$\\
            $D$ (Mpc) & $3.86\pm 0.05^{\hspace{0.03cm} b}$\\
            $\mu_{r',0} (\mathrm{mag \hspace{0.05cm} arcsec^{-2}})$ & 22.6$^{\hspace{0.03cm} c}$\\
		$\mu_{r',e} (\mathrm{mag \hspace{0.05cm} arcsec^{-2}})$ & 24.0$^{\hspace{0.03cm} c}$\\
		$r_e (\mathrm{kpc})$ & 2.7$^{\hspace{0.03cm} c}$\\
            $m_V$ & 10.87$\pm0.03^{\hspace{0.03cm} d}$\\
            $M_V$ & $-17.06 \pm 0.04$\\
            Metallicity ($Z_\odot$) & 0.3$^{\hspace{0.03cm} e}$\\
            Inclination (deg) & 63$^{\hspace{0.03cm} f}$\\
            PA of major axis (deg) & 55$^{\hspace{0.03cm} f}$\\
            log$M_* \hspace{0.05cm} (M_{\odot})$ & 8.39$^{\hspace{0.03cm} g}$\\
            $E(B-V)$ & 0.036$^{\hspace{0.03cm} h}$\\
            $v_{\mathrm{max}}$ (km s$^{-1}$) & $77.6\pm 5.0^{\hspace{0.03cm} i}$\\
            log$M_{200} \hspace{0.05cm} (M_{\odot})$ & $10.93\pm0.08$\\ 
            $N_{\mathrm{GC}}$ & $27 \pm 5$\\
            $S_N$ & $4.0 \pm 0.8$\\
		\hline
            \hline
	\end{tabular}
\end{table}

\section{Data and Methodology}
\label{sec:data&method}

\subsection{Data}
\label{subsec:data}

We search for globular clusters in IC~2574 using public HST/ACS imaging from the Mikulski Archive for Space Telescopes (MAST) Portal from the Space Telescope Science Institute (STScI) in the F814W and F555W broadband filters \citep{Fleming_MAST_2015}. Candidate GC systems have been similarly examined in HST/ACS imaging, ranging across nearby giant elliptical galaxies, spiral galaxies, dwarf spheroidals (dSph) and dIrrs, and distant unbound GCs in galaxy clusters \citep{Forbes_1997,Chandar_2004,Sharina_2005,Hwang_Lee_2006,Barmby_2006,Nantais_2010,Santiago_Cort_s_2010,West_2011,Simanton_2015,LomeliNunez_2022,Hazra_2022}, as well as with similar methods on imaging from other telescopes \citep{Goudfrooij_2003,Tudorica_2015,Gonzalez-Lopezlira_2017,Cantiello_2018}.\\

\begin{figure}
    \centering
    \includegraphics[width=\columnwidth]{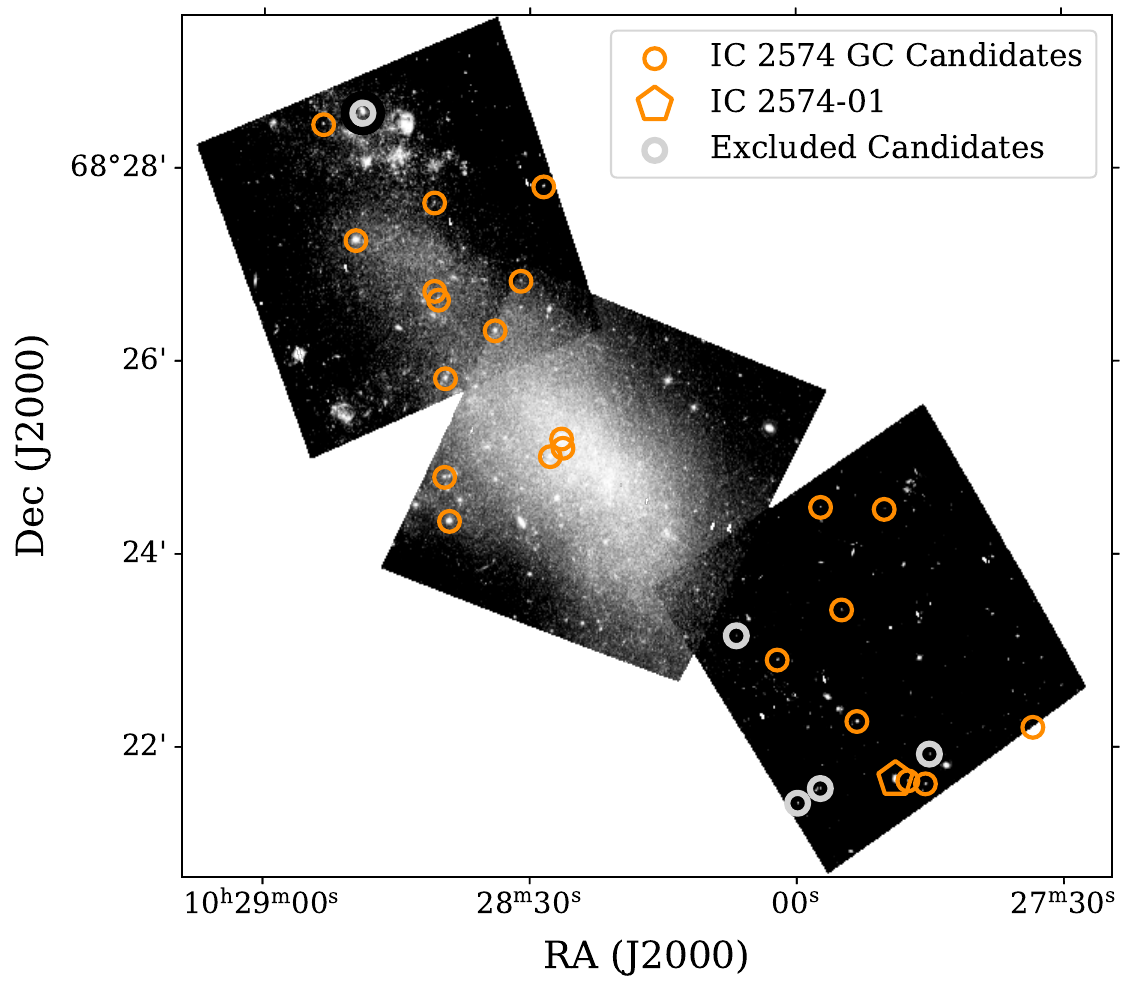}
    \caption{Mosaic of IC~2574 pointings with the distribution of GC candidates overlaid in orange circles (highest luminosity candidate IC 2574-01 is an orange pentagon). In light grey circles we have 5 excluded candidates that we speculate could be young star clusters, not GCs. From left to right we label the pointings: 9755 (East), 10605-1 (Central), 10605-2 (West), as North is up and East is to the left. \citet{Weisz_2008} found that young blue stars are distributed across the galaxy (including the southern part of the West field), with the highest concentration in the northern part of the East field. Old red stars are elliptically distributed with a major axis in the northeast-southwest direction, with the highest concentration in the overlap region between the East and Central field.}
    \label{fig:ic2574map}
\end{figure}

IC~2574 is visible over three pointings that each cover a $202\times202$ arcsec field-of-view (FOV), depicted in a mosaic in Fig. \ref{fig:ic2574map} with the GC candidates shown in orange. The left pointing was obtained from program 9755 \citep[PI:~][]{Walter_HST_9755}, and the central and right pointings from program 10605 are referred to in this paper as 10605-1 and 10605-2 respectively \citep[PI:~][]{Skillman_HST_10605}. Table \ref{tab:observationlog} outlines the observation details.

\begin{table}
	\centering
	\caption{Observation log details for the three pointings; (a) \citet{Walter_HST_9755}, (b) \citet{Skillman_HST_10605}.}
	\label{tab:observationlog}
	\begin{tabular}{llll}
		\hline
            \hline
             & \vtop{\hbox{\strut 9755$\hspace{0.03cm}^a$}\hbox{\strut [East]}} & \vtop{\hbox{\strut 10605-1$\hspace{0.03cm}^b$}\hbox{\strut [Central]}} & \vtop{\hbox{\strut 10605-2$\hspace{0.03cm}^b$}\hbox{\strut [West]}}\\
		\hline
            \hline
             \\
            Name & IC2574-SGS & IC-2574-1-COPY & IC-2574-2\\
            RA, Dec (J2000) & \vtop{\hbox{\strut 10:28:43.17,}\hbox{\strut +68:27:04.56}} & \vtop{\hbox{\strut 10:28:23.07,}\hbox{\strut +68:24:36.00}} & \vtop{\hbox{\strut 10:27:50.00,}\hbox{\strut +68:22:54.99}}\\
            Observation Date & 2004-02-06 & 2006-03-20 & 2006-01-27\\ 
            Exposure Time (s) & 6400.0 & 4784.0 & 4784.0\\
             \\
		\hline
            \hline
	\end{tabular}
\end{table}

\subsection{Visual Inspection}
\label{subsec:visual}

In the absence of matched filter techniques or machine learning algorithms specifically designed for the identification of GCs, visual inspection of imaging data remains a common technique for identifying these objects \citep[][and references therein]{Hwang_Lee_2006,Barmby_2006,Nantais_2010}. Visual inspection involves scrutinising the images to identify the distinct sources that exhibit the characteristics indicative of GCs.
As IC~2574 is relatively nearby, its GCs are resolved in size, but the individual stars are not resolved. This means they appear as compact and round partially extended objects, where only their outermost stars can be resolved around the full edge of the clusters. This makes their appearance discernible from individual stars, clusters of stars or galaxies. Visual inspection of the candidates does impart subjectivity to the selection procedure that cannot have a well-defined completeness estimate, but detailed analysis of the individually selected candidates aids in the final selection \citep{Barmby_2006,Nantais_2010}.

We use SAOImage DS9 \citep{DS9_2000,Joye_DS9_2003} to visually inspect the images in the deeper F814W filter to find circular objects that have stars resolved around the edges, and that importantly maintain their circular shape across varying contrasts; objects found to have non-circular shapes at extreme contrasts are dismissed from consideration. We scan across all areas of each pointing image in DS9 to visually select 28 candidates based on these distinctive features. The next step is to obtain the photometry for these candidates in order to check whether their size and colour are in line with those expected for GCs.

\subsection{Photometry}
\label{subsec:photometry}

To evalulate the photometry of our selected candidates, we perform source detection and aperture photometry using \texttt{Source Extractor} (\texttt{SE}) \citep{Bertin_Arnouts_1996}. We use \texttt{SE} in dual image mode with the F814W filter as the detection image, cycling through with the F814W and F555W filters as the measurement images. Using the input parameters in Table \ref{tab:inputparams}, as in \citet{LomeliNunez_2022} and following techniques from \citet{Holwerda_2005}, we run \texttt{SE} from the command line, selecting the output parameters \texttt{FLUX\_BEST}, which is useful for well-resolved sources that do not have a complex morphology, the J2000 coordinates \texttt{ALPHA\_J2000} and \texttt{DELTA\_J2000} and the \texttt{FLUX\_RADIUS} set to 0.5, which corresponds to the half-light radius ($r_h$) of the source\footnote{When selecting \texttt{FLUX\_RADIUS} in the \texttt{``-.param"} file, comment out the description line that follows it in order to avoid a syntax error.}.

Each pointing has varying levels of crowding, therefore it is important to account for background subtraction in a more accurate way unique to each source. Setting \texttt{BACKPHOTO\_TYPE} to LOCAL calculates a local background estimation in an annulus around each source, which is far more accurate in this context than otherwise using the GLOBAL background subtraction that consists of an average value across the entire image, leading to inaccurate photometry.
 
 \begin{table}
	\centering
	\caption{The input parameters for the Source Extractor runs.}
	\label{tab:inputparams}
	\begin{tabular}{ll}
		\hline
            \hline
		Input Parameter &  \\
		\hline
            \hline
		\texttt{DETECT\_MINAREA} & 5\\
		\texttt{DETECT\_THRESH} & 1.4\\
		\texttt{BACK\_SIZE} & 32\\
		\texttt{BACK\_FILTERSIZE} & 3\\
		\texttt{PIXEL\_SCALE} & 0.05\\
		\texttt{PHOT\_FLUXFRAC} & 0.5\\  
		\texttt{BACKPHOTO\_TYPE} & LOCAL\\
		\hline
            \hline
	\end{tabular}
\end{table}

For the extinction correction of the fluxes, we use the extinction ratios $A(P)~/~E(B-V)$ of 1.825 and 3.177 for F814W and F555W respectively, as outlined in Table 14 of \citet{Sirianni_2005}; we use the spectral energy distribution G2, which is an appropriate choice for HST/ACS photometry because they are similar to that of the typical stars we find in GCs, as in e.g. \citet{Galleti_2006}. We take the extinction value $E(B-V)=0.036$ for IC~2574 from the NASA/IPAC Infrared Science Archive \citep{SFD_1998}, which gives us $A_{\mathrm{F814W}}$ = 0.0657 and $A_{\mathrm{F555W}}$ = 0.114. To convert the fluxes to apparent magnitudes, we use the zeropoints ($ZP$) for the ACS Wide Field Channel (WFC) in ABMAG in Table 10 of \citet{Sirianni_2005} in the magnitude equation 

\begin{equation}
   m_{\rm filter}= -2.5 \times \mathrm{log}_{10}(\mathrm{flux}) + ZP\hspace{0.025cm};
\end{equation}
\vspace{0.05cm}

\noindent where $ZP_{\mathrm{F814W}}$ = 25.937 and $ZP_{\mathrm{F555W}}$ = 25.718. To then convert the apparent magnitudes to absolute magnitudes, we use the IC~2574 dereddened distance modulus of $27.93\pm0.03$ mag \citep{Dalcanton_2009}.

For the GC selection criteria, colour cuts have been performed in the filters V and I \citep[e.g.][]{Forbes_1997,Goudfrooij_2003,Chandar_2004,Sharina_2005,West_2011,Simanton_2015}, so we perform standard filter transformations from our WFC filters on the HST/ACS to the Johnson-Cousins BVRI system as outlined in \citet{Sirianni_2005} using the coefficients in Table 18 in the `observed' panel. The observed coefficients are determined from observations of standard stars and tend to be more reliable for accurate photometry in the WFC and BVRI filter systems than the synthetic coefficients. This filter transformation is performed using Equation 12 in \citet{Sirianni_2005}.

\subsection{Globular Cluster Selection}
\label{subsec:selection}

For the final selection of our GC candidates, we apply broad cuts to our dataset in size and colour.
We adopt a generous colour-cut of $0.3~\leq~V-I~\leq~2.0$ as in \citet{Goudfrooij_2003}.
The HST/ACS pixel scale is 0.05 arcsec/pixel, so the projected spatial resolution for the IC~2574 distance (3.86 Mpc) is 0.935 pc/pixel. Objects found to have sizes below $\sim$1 pc were dismissed as likely point sources (stars) based on the resolution of HST/ACS, increasing the likelihood of identifying genuine GCs and reducing the contamination of false positives caused by these individual stars or compact clusters of stars; above this 1 pc size, all GC candidates are resolved in size. Our smallest candidate is $\sim$1.4 pc, well above this 1 pc lower limit we have defined.
We expect some candidates are significantly more extended than others, so we set our upper limit size cut to an extremely extended 30 pc, consistent with extended clusters seen in Andromeda \citep{Huxor_2005}.

After applying these cuts, we find 23 GC candidates, which need to be corrected for observational incompleteness.
The 5 candidates that fall outside of the cuts are referred to from here on as the `excluded candidates'; these excluded candidates have appropriate sizes but are too blue in colour to be considered GC candidates.

\subsection{Completeness Corrections}
\label{subsec:completeness}
There are two sources of incompleteness that we must consider in our analysis; the radial distribution and the magnitudes of the GC candidates.\\

\begin{figure}
    \centering
    \includegraphics[width=\columnwidth]{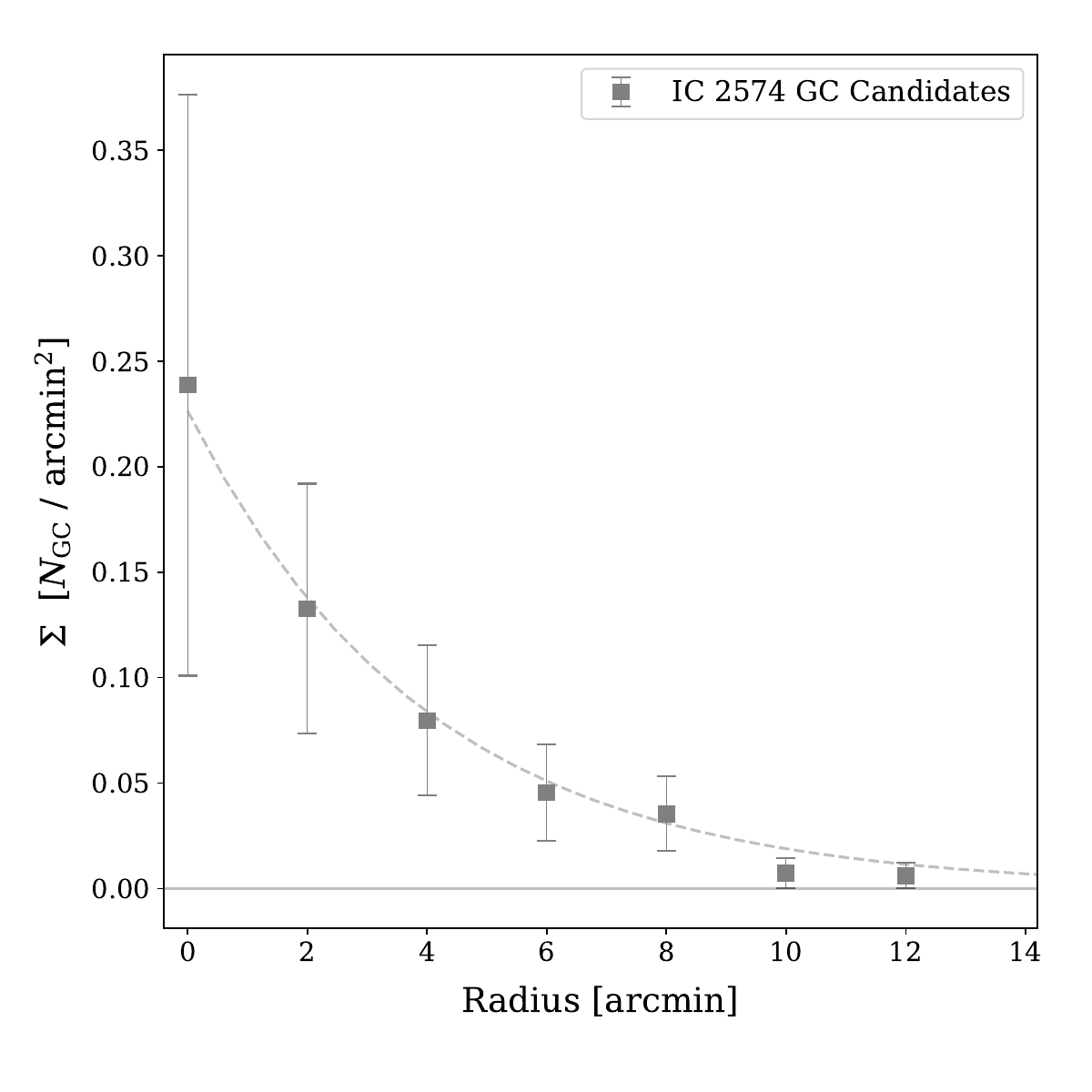}
    \caption{The radial profile of the candidate GCs of IC~2574, which shows the number of GCs per unit area ($\Sigma$) as a function of the distance from the centre in arcmin. Error bars in $\Sigma$ are Poisson uncertainties. We have candidates out to 12.8 arcmin. The curve fit is applied only to the first 3.37 arcmin of data as this is the extent of our radial completeness in the minor axis direcion, which is one pointing across (ACS FOV width). We extend the curve fit out to 30 arcmin to approximate where it approaches $\Sigma=0 \hspace{0.05cm} \mathrm{arcmin}^{-2}$ (solid grey line). The ratio between the integrals under the curve from 0 to 12.8 arcmin and 12.8 to 30 arcmin are used to calculate the radial completeness correction (see text for details).}
    \label{fig:radialprofile}
\end{figure}

We first consider the radial distribution of the GC candidates around IC~2574. In Fig. \ref{fig:radialprofile} we plot the radial profile of the GC candidates, i.e. $N_{\mathrm{GC}}$ per unit area ($\Sigma$) as a function of their distance from the centre of the galaxy (see Table \ref{tab:ic2574properties} for central coordinates). From Fig. \ref{fig:ic2574map} we can see that there is good coverage along the major axis, however the incompleteness arises in the direction of the minor axis as this is only covered by one pointing across. We perform curve fitting of the GC radial profile from 0 to 3.37 arcmin (ACS FOV: $202 \times 202$ arcsec) as this is the extent of the completeness in the minor axis direction. We use the method outlined in \citet{Martin_2016}, where we assume that our GC system's radial distribution follows the same behaviour as would be expected for the stellar distribution of a galaxy:

\begin{equation}
   \rho_{\rm dwarf} (r)=\frac{1.68^2}{2\pi R_{\rm GC}^2(1-\epsilon)}N_{\rm GC}\exp{\left[\frac{-1.68r}{R_{\rm GC}}\right]},
    \label{eq:densityprofile}
\end{equation}

\noindent where we substitute $N_*$ and $r_h$ in the literature for our current value of $N_{\mathrm{GC}}$ = 23 and fit for $R_{\rm GC}$ which is a scale radius that tells us the radius at which half the number of GCs per arcmin$^2$ are found, and where $r$ is the elliptical radius such that: 

\begin{equation}
    \begin{split}
        r = \Bigg[ \hspace{0.2cm} \Big(\frac{1}{1-\epsilon}((x-x_{0})\cos{\theta}-(y-y_{0})\sin{\theta})\Big)^2 \hspace{0.1cm} + \\
        \hspace{-0.5cm} \Big((x-x_{0})\sin{\theta}-(y-y_{0})\cos{\theta}\Big)^2 \hspace{0.2cm} \Bigg] ^{\frac{1}{2}}.
    \label{eq:ellipticalradii}
    \end{split}
\end{equation}

\noindent In this equation, $\theta$ is the position angle (PA) of the major axis, $x_{0}$ and $y_{0}$ are the central coordinates of IC~2574 and $x$ and $y$ are the coordinates of the GC candidate. We assume that the GCs tend towards a spherical distribution, and set the ellipticity $\epsilon$ to 0.

This curve fit agrees exceptionally well with the GC radial profile and shows the farthest GCs almost reaching $\Sigma= 0 \hspace{0.06cm} \mathrm{arcmin}^{-2}$, which indicates that the radial completeness corrections are not substantial. In order to apply the corrections, we integrate under the curve from 0 until the maximum radial distance of GCs (12.8 arcmin), and then from this maximum to a distant 30 arcmin, where we approximate $\Sigma=0 \hspace{0.06cm} \mathrm{arcmin}^{-2}$ has been approached. We obtain the ratio between these two integrals to obtain the radial completeness correction, which we factor in with the magnitude completeness correction below. 

We perform completeness corrections in magnitude using the GC luminosity function\footnote{See Sec.~\ref{subsubsec:gclf} for a description of the GCLF.} (GCLF), as in e.g. \citet{Richtler_2003_GCLF,Beasley_2020}. We use the more luminous half of the GCLF, above the peak, and reflect this side to create a mirrored distribution that accounts for low-luminosity GCs that may have been missed, especially in the crowded central region of IC~2574. We fit a Gaussian distribution to the original GCLF and the mirrored GCLF, and integrate under the curve for each, using the ratio between these integrals to multiply and scale up $N_{\mathrm{GC}}$.\\

Accounting for both the radial and magnitude completeness corrections gives us a completeness-corrected value of $N_{\mathrm{GC}}~=~27~\pm~5$, including its Poissonian uncertainty, which we use for all subsequent analysis.

\section{Results and Discussion}
\label{sec:results}

\subsection{The GC System}
\label{subsec:GCsystem}

\begin{figure*}
    \centering
    \includegraphics[width=\textwidth]{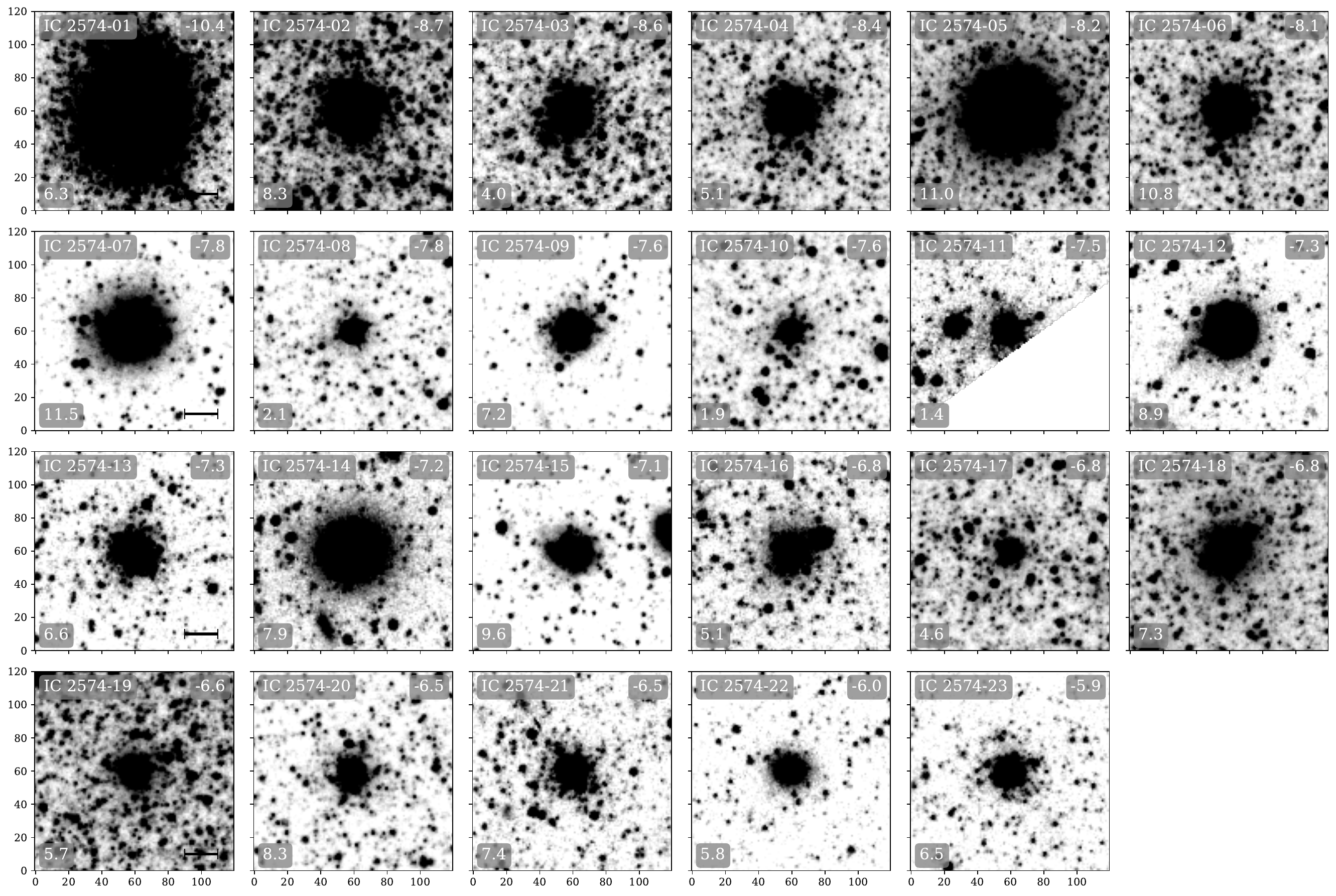}
    \caption{Globular cluster candidates of IC~2574 ordered by $M_V$. All cutout images are centred on the candidates in a $120\times120$ pixel box, with the GC candidate name in the top left, $M_V$ in the top right and $r_h$ in the bottom left in pc. The bar in the bottom right of the first column of cutouts indicates 1" in the image. North is up, East is to the left.}
    \label{fig:finalGCcandidates}
\end{figure*}

\begin{figure*}
    \centering
    \includegraphics[width=\textwidth]{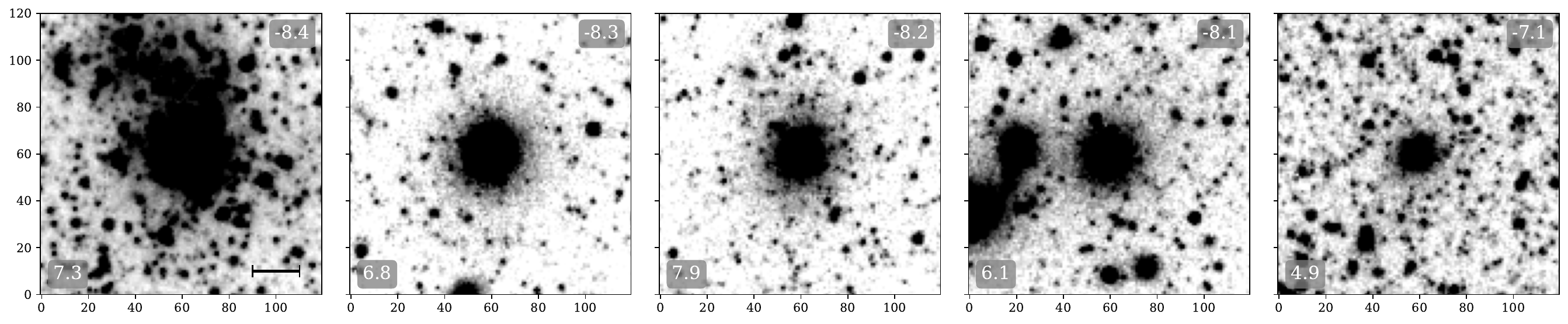}
    \caption{The 5 excluded candidates that do not pass the cuts outlined in Sec. \ref{subsec:selection}, ordered by $M_V$. All cutout images are centred on the object in a $120\times120$ pixel box, with $M_V$ in the top right and $r_h$ in the bottom left in pc. The bar in the bottom right of the first cutout indicates 1" in the image. These 5 objects are also shown in Fig.~\ref{fig:Mvsize} and Fig.~\ref{fig:colourdistribution} as light grey data points; these are dismissed from the final GC candidate selection as although they are within the size cut, they fall beyond the colour cut with extra blue colour. We speculate that these could be young star clusters given their extremely blue colour. North is up, East is to the left.
    }
    \label{fig:example_notincut}
\end{figure*}

A compilation of the cutouts for all 23 GC candidates identified are shown in Fig.~\ref{fig:finalGCcandidates}.
Our candidate GCs have luminosities spanning the range $-5.9~\geq~M_V~\geq~-10.4$. Our highest luminosity GC candidate (IC~2574-01) has $M_V = - 10.4$, comparable to $\omega$ Cen \citep{Harris_2010}. The sizes of our GC candidates range from $1.4 \leq r_h \leq 11.5$ pc, with colours in the range $0.52 \leq V-I \leq 1.59$, typical of GCs \citep[e.g.][]{Forbes_1997,Goudfrooij_2003,Chandar_2004,Sharina_2005,Santiago_Cort_s_2010,LomeliNunez_2022}. Our most extended GC candidate (IC~2574-07) has a size of 11.5~pc.
The excluded candidates that were in the initial visual selection fell outside of the broad cuts because of their extremely blue ($V-I <-0.5$) colour, these are shown in Fig.~\ref{fig:example_notincut}. These objects can be seen to have similar morphological characteristics to our final GC candidates in Fig.~\ref{fig:finalGCcandidates}.\\

\begin{figure}
    \centering
    \includegraphics[width=\columnwidth]{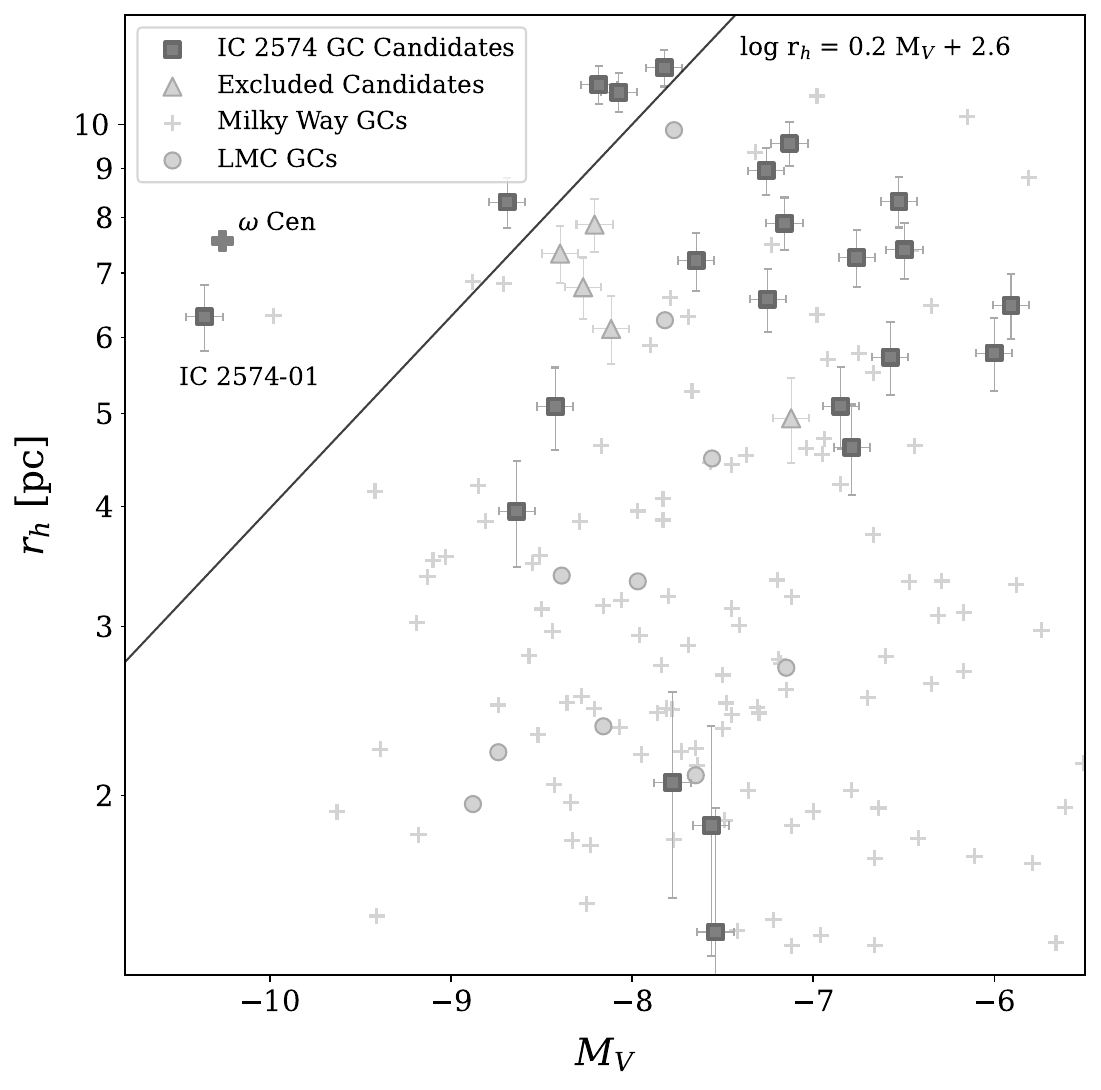}
    \caption{The relation between the size $r_h$ (pc) and luminosity $M_V$ of our GC candidates. The GC candidates do not spread across the $r_h$ vs. $M_V$ plane uniformly but tend to abide by the relation log $r_h = 0.2 \hspace{0.05cm} M_V + 2.6$ (solid line), that sets an envelope that is typically traced by GCs \citep{vandenBergh_2004,Mackey_2013}. This relation illustrates that larger GCs are intrinsically less luminous, more luminous GCs tend towards being more compact, and typically GCs do not rise above this envelope. The MW GC $\omega$ Cen is a known exception to this rule (lying above the boundary), which is plotted on this distribution (labelled bold plus symbol) with a more recent $r_h$ value from \citet{Evans_2022}. All of the other MW GCs from the \citet{Harris_2010} catalogue are plotted as light grey plus symbols. We also plot selected LMC GCs from \citet{McLaughlin_2005} as light grey circles for a dwarf system comparison. IC~2574-01 is also a clear exception to this rule, situated close to $\omega$ Cen, well beyond the boundary of the envelope; it is plausible that our $\omega$ Cen-like candidate is similar in nature. We also have 4 other GC candidates that lie just above the relation, these further exceptions could suggest a unique story about the formation of IC~2574's GC system.}
    \label{fig:Mvsize}
\end{figure}

We plot the well-known relation between the size and luminosity of our GC candidates in Fig. \ref{fig:Mvsize}, with GCs from the MW and the dIrr Large Magellanic Cloud (LMC) plotted for comparison \citep{Harris_2010,McLaughlin_2005}. We overplot the relationship log $r_h = 0.2 \hspace{0.05cm} M_V + 2.6$ from \citet{vandenBergh_2004}, depicting a characteristic envelope typically traced by MW GCs \citep{Mackey_2013}.
This relation illustrates that larger GCs are intrinsically less luminous, more luminous GCs tend towards being more compact, and typically, few GCs rise above this envelope.
An exception to this rule is found by \citet{vandenBergh_2004} in the GC NGC~2419, who suspect that it could be the surviving core of a dSph galaxy, much like suggestions that $\omega$ Cen is also the surviving core of a disrupted dwarf galaxy \citep[][and references therein]{Majewski_2000,Hilker_2000,Bekki_2003,Youakim_2023}. We find that IC~2574-01 is in close proximity to $\omega$ Cen on this plane, suggesting it could be similar in nature to $\omega$ Cen. We also have 4 other GC candidates that lie just above the relation, which is intriguing. Interestingly, the GCs of IC~2574 are on average larger than those of the MW and LMC. Several works have shown that dwarfs, such as ScI-dE1, Fornax, NGC~6822, IKN, and Sagittarius, may host more extended clusters owing to a weaker tidal field than higher mass galaxies \citep{DaCosta_Extended_2009,Georgiev_2009,Harris_2010,Hwang_2011,Cole_2012,Veljanoski_2015,Tudorica_2015,Orkney_2019}. IC~2574's GC system could contribute to this valuable literature if we can confirm the GCs as members of IC~2574 with spectroscopy and make a statistically significant comparison.

\begin{figure}
    \centering
    \includegraphics[width=\columnwidth]{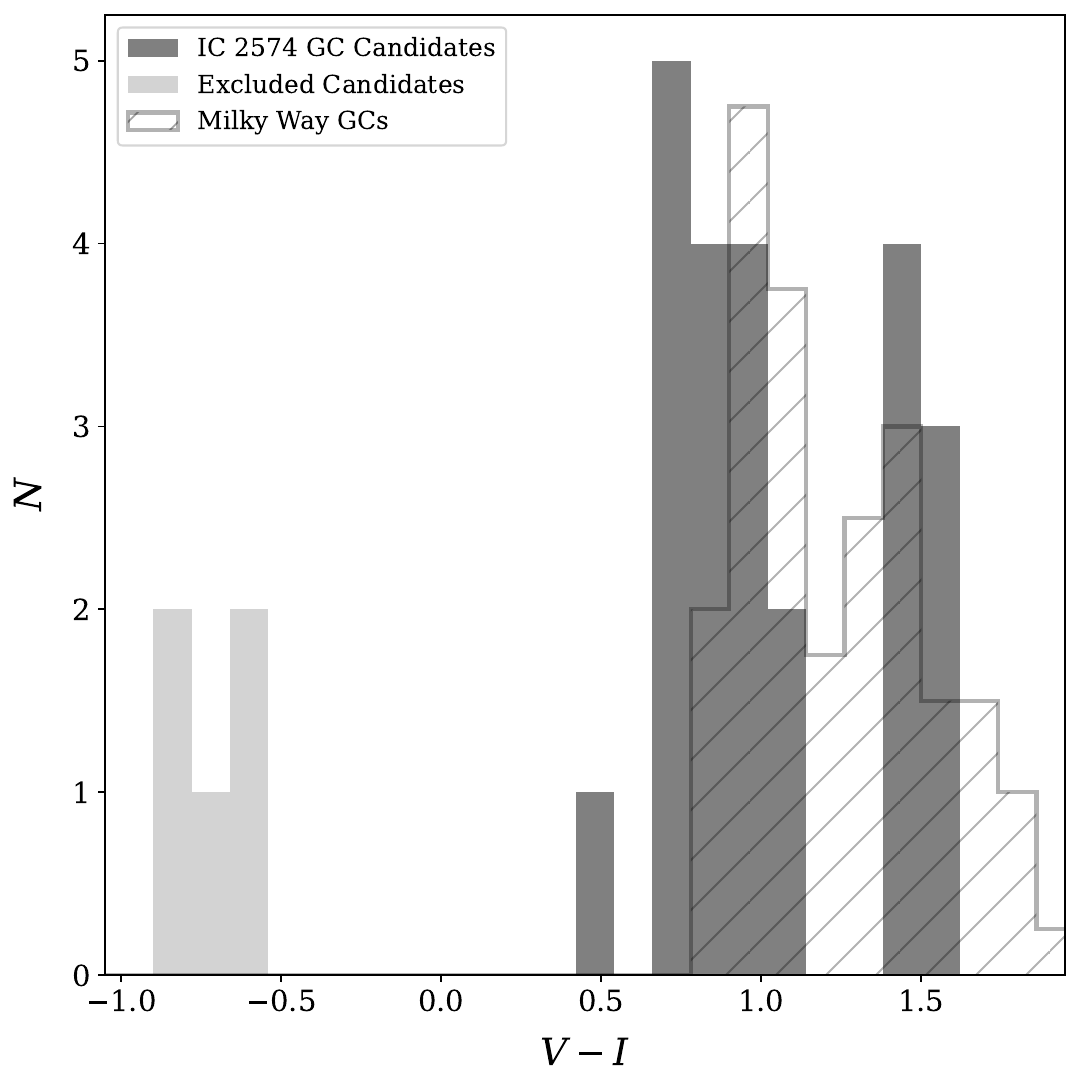}
    \caption{The colour distribution of the 23 candidate GCs of IC~2574 (dark grey), which appears to exhibit a bimodal feature. We plot the $V-I$ colour distribution of the MW GCs \citep{Harris_2010} with a height scaled down by a factor of 1/4 to emphasise the apparent colour bimodality of the IC~2574 GC candidates in comparison to the MW GCs. On the far left with the extremely blue colours are the 5 excluded candidates that fall outside of the $0.3 \leq V-I \leq 2.0$ colour cut (light grey), which could potentially be young star clusters linked to the recent star formation that occurred in IC~2574 $\sim$1~Gyr ago.}
    \label{fig:colourdistribution}
\end{figure}

The 5 light grey data points (triangle symbols) in the size-luminosity plot in Fig.~\ref{fig:Mvsize} are the excluded candidates. They have appropriate sizes as can be seen in this distribution, however the colour distribution in Fig.~\ref{fig:colourdistribution} shows that they have extremely blue colours ($V-I<-0.5$), well below the generous lower boundary of the colour cut ($V-I=0.3$). It is possible that these 5 excluded candidates could be young massive star clusters, because of their morphology and the extremely blue colour that they exhibit \citep{PortegiesZwart_2010}. Young massive star clusters are known to form in a variety of starburst environments \citep{Meurer_1995, Watson_1996}, consistent with the recent intense burst of star formation that occurred in IC~2574 $\sim$1~Gyr ago
\citep{Weisz_2008}. The recent burst accounts for 15\% of the stars formed, which may be consistent with the fraction of excluded candidates (potential young star clusters) to all candidate clusters found, which is $\sim 16\%$.

\begin{figure}
    \centering
    \includegraphics[width=\columnwidth]{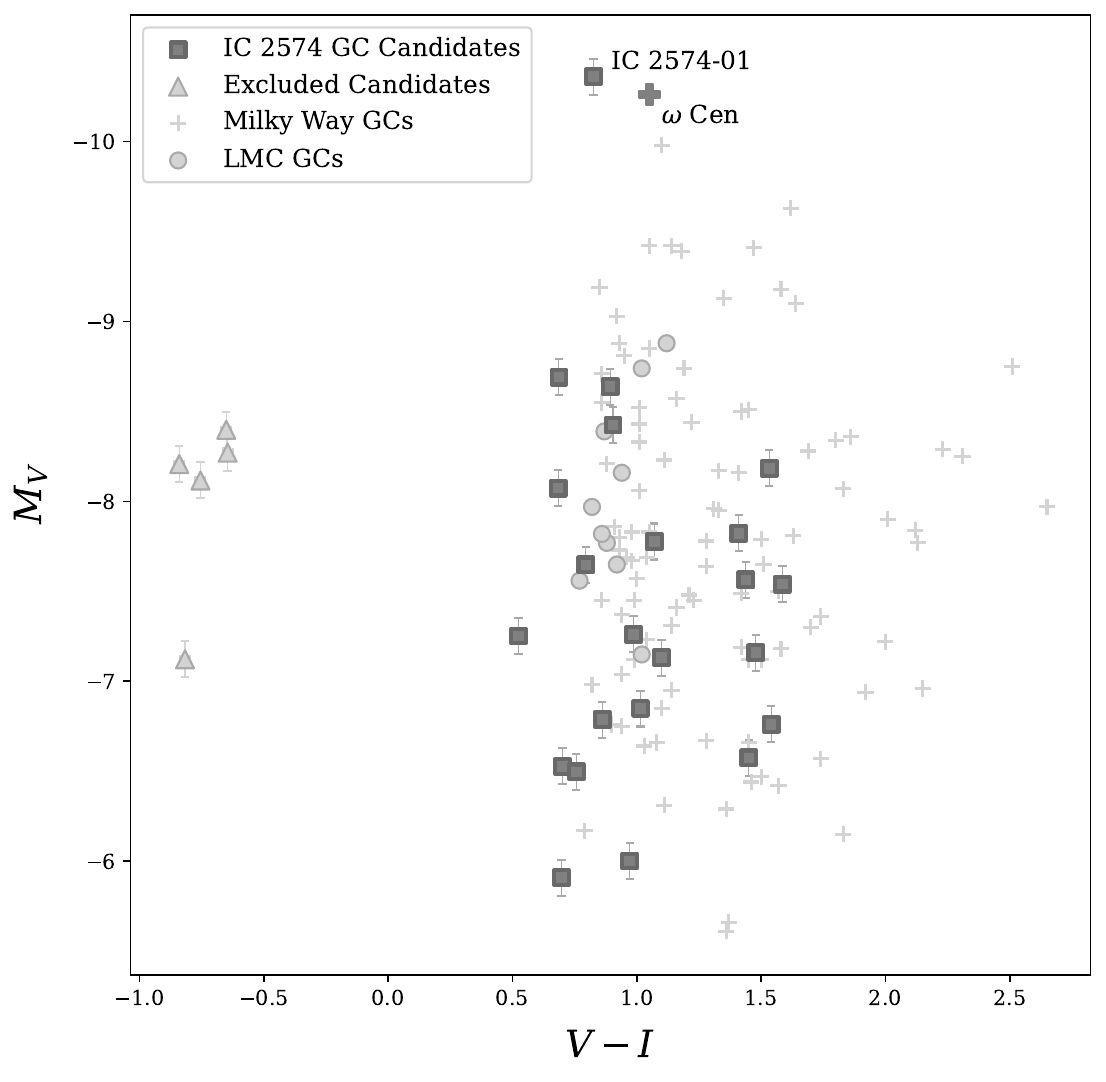}
    \caption{The colour-magnitude diagram, with $V-I$ vs. $M_V$, for our GC candidates and excluded candidates. We plot the Milky Way GCs \citep{Harris_2010} and selected LMC GCs \citep{McLaughlin_2005} for comparison.}
    \label{fig:CMD}
\end{figure}

\begin{table*}
	\centering
	\caption{A catalogue of the GC candidate properties: the RA, Dec coordinates in J2000, the $V-I$ colour, the absolute magnitude $M_V$, the half-light radius $r_h$ in pc and its field. All $r_h$ values are assumed to have an uncertainty of 0.5 pc, accounting for the limitations in instrument resolution. The uncertainties for all colour values are 0.02 mag and the uncertainties for the $M_V$ values are 0.1 mag; the uncertainties are propagated through from the uncertainties in the flux obtained from \texttt{Source Extractor}, the filter transformations and the distance to IC~2574. The bottom section of the table has the same information for the `excluded' candidates that could be young massive star clusters (YMC).}
	\label{tab:GCcandidates}
	\begin{tabular}{llllllllllll}
		\hline
            \hline
            Name &   & RA (J2000) &   & Dec (J2000) &   & V-I &   & $M_V$ &   & $r_h$ (pc) & Field\\
            \hline
            \hline
             \\
            IC 2574-01 &   & 10:27:48.90 &   & +68:21:40.30 &   & 0.83 &   & -10.4 &   & 6.3 & 10605-2\\
            IC 2574-02 &   & 10:28:26.44 &   & +68:25:11.86 &   & 0.69 &   & -8.7 &   & 8.3 & 10605-1\\
            IC 2574-03 &   & 10:28:27.70 &   & +68:25:00.86 &   & 0.89 &   & -8.6 &   & 4.0 & 10605-1\\
            IC 2574-04 &   & 10:28:33.93 &   & +68:26:19.05 &   & 0.90 &   & -8.4 &   & 5.1 & 9755\\
            IC 2574-05 &   & 10:28:49.64 &   & +68:27:15.01 &   & 1.53 &   & -8.2 &   & 11.0 & 9755\\
            IC 2574-06 &   & 10:28:40.32 &   & +68:26:38.07 &   & 0.68 &   & -8.1 &   & 10.8 & 9755\\
            IC 2574-07 &   & 10:28:39.05 &   & +68:24:20.59 &   & 1.41 &   & -7.8 &   & 11.5 & 10605-1\\
            IC 2574-08 &   & 10:28:31.02 &   & +68:26:49.89 &   & 1.07 &   & -7.8 &   & 2.1 & 9755\\
            IC 2574-09 &   & 10:28:28.46 &   & +68:27:48.53 &   & 0.80 &   & -7.6 &   & 7.2 & 9755\\
            IC 2574-10 &   & 10:28:40.78 &   & +68:27:38.47 &   & 1.44 &   & -7.6 &   & 1.9 & 9755\\
            IC 2574-11 &   & 10:27:33.44 &   & +68:22:12.26 &   & 1.59 &   & -7.5 &   & 1.4 & 10605-2\\
            IC 2574-12 &   & 10:27:45.53 &   & +68:21:37.29 &   & 0.99 &   & -7.3 &   & 8.9 & 10605-2\\
            IC 2574-13 &   & 10:27:47.50 &   & +68:21:39.19 &   & 0.52 &   & -7.3 &   & 6.6 & 10605-2\\
            IC 2574-14 &   & 10:27:53.22 &   & +68:22:16.14 &   & 1.48 &   & -7.2 &   & 7.9 & 10605-2\\
            IC 2574-15 &   & 10:28:39.61 &   & +68:24:47.96 &   & 1.10 &   & -7.1 &   & 9.6 & 10605-1\\
            IC 2574-16 &   & 10:28:02.17 &   & +68:22:54.39 &   & 1.02 &   & -6.8 &   & 5.1 & 10605-2\\
            IC 2574-17 &   & 10:28:40.75 &   & +68:26:43.55 &   & 0.86 &   & -6.8 &   & 4.6 & 9755\\
            IC 2574-18 &   & 10:28:39.52 &   & +68:25:49.22 &   & 1.54 &   & -6.8 &   & 7.3 & 9755\\
            IC 2574-19 &   & 10:28:26.28 &   & +68:25:06.09 &   & 1.45 &   & -6.6 &   & 5.7 & 10605-1\\
            IC 2574-20 &   & 10:28:53.32 &   & +68:28:27.01 &   & 0.70 &   & -6.5 &   & 8.3 & 9755\\
            IC 2574-21 &   & 10:27:54.92 &   & +68:23:25.49 &   & 0.76 &   & -6.5 &   & 7.4 & 10605-2\\
            IC 2574-22 &   & 10:27:50.10 &   & +68:24:27.95 &   & 0.97 &   & -6.0 &   & 5.8 & 10605-2\\
            IC 2574-23 &   & 10:27:57.27 &   & +68:24:29.27 &   & 0.70 &   & -5.9 &   & 6.5 & 10605-2\\
             \\
            \hline
            \hline
             \\
            IC 2574-YMC-01 &   & 10:28:48.90 &   & +68:28:34.30 &   & -0.65 &   & -8.4 &   & 7.3 & 9755\\
            IC 2574-YMC-02 &   & 10:27:59.89 &   & +68:21:25.39 &   & -0.65 &   & -8.3 &   & 6.8 & 10605-2\\
            IC 2574-YMC-03 &   & 10:27:45.07 &   & +68:21:55.79 &   & -0.84 &   & -8.2 &   & 7.9 & 10605-2\\
            IC 2574-YMC-04 &   & 10:27:57.34 &   & +68:21:34.50 &   & -0.75 &   & -8.1 &   & 6.1 & 10605-2\\
            IC 2574-YMC-05 &   & 10:28:06.76 &   & +68:23:09.50 &   & -0.82 &   & -7.1 &   & 4.9 & 10605-2\\
             \\
            \hline
            \hline
	\end{tabular}
\end{table*}

\subsection{Colour Bimodality in the GC System of IC~2574}
\label{subsubsec:bimodality}

The colour distribution of the 23 GC candidates, as well as the 5 excluded candidates, is shown in Fig. \ref{fig:colourdistribution}. For the IC~2574 GC candidates (dark grey) there appears to be a bimodal feature, which we fit with a double Gaussian. We find 30\% localised around the redder $V-I=1.5\pm0.1$ ($\sigma=0.1\pm0.1$) and 70\% around the bluer $V-I=0.84\pm0.05$ ($\sigma=0.18\pm0.05$).
The redder GC candidates are spatially distributed across the full major axis of the galaxy, which reassures us that their redder colour is an intrinsic property and is likely not a systematic reddening effect (implied if they were centrally located). We also plot the $V-I$ bimodal colour distribution of the MW GCs \citep{Harris_2010} in Fig.~\ref{fig:colourdistribution} to emphasise the bimodal feature exhibited by IC~2574. We scaled down the histogram height of the MW GCs by a factor of 4 to make a sensible visual comparison.

Colour bimodality is a significant characteristic of a GC system \citep{Zepf_1993,Gebhardt_1999,Larsen_2001}, understood to be a proxy for an underlying bimodality in metallicity \citep{KisslerPatig_1998,Brodie_2006,ChiesSantos_2012,Fahrion_2020,Harris_2023,Hartman_2023}.
A metallicity bimodality is often interpreted as a signature of distinct formation channels within a GC system \citep{Harris_1994,Meurer_1995,Watson_1996,Forbes_1997,Cote_1999_MW,Larsen_2001,Brodie_2006,Kruijssen_2014,Choksi_2019,ReinaCampos_Formation_2022}, e.g. those formed in-situ and those accreted onto the galaxy \citep{Zepf_1993,Forbes_1997,Gebhardt_1999,Larsen_2001,Kundu_2001,Peng_2006,Brodie_2006,LomeliNunez_2022}.
The MW and M81 GC systems are examples of galaxies exhibiting bimodal colour and confirmed intrinsic bimodal metallicity distributions \citep{Cote_1999_MW,Perrett_2002_MW_M31,Chandar_2001_M81,Ma_2005,Nantais_2010,Santiago_Cort_s_2010,Nantais_2011}.

As such, IC~2574's bimodal colour distribution could suggest two subpopulations with different formation mechanisms or alternatively an in-situ population and an accreted metal-poor GC population from a lower mass galaxy. Although IC~2574 has no signs of present interaction, there is the possibility of a past merger as dwarf-dwarf mergers have been observed \citep[e.g.][]{Rich_2012,MartinezDelgado_2012}, simulated and discussed in the literature \citep{Forbes_2010,Deason_DwarfMergers_2014,Wetzel_2015,Wheeler_SoS_2015,Deason_2015}. The presence of two different metallicity populations would require confirmation with spectroscopy of the IC~2574 GCs.

\subsection{The GC Luminosity Function}
\label{subsubsec:gclf}

\begin{figure}
    \centering
    \includegraphics[width=\columnwidth]{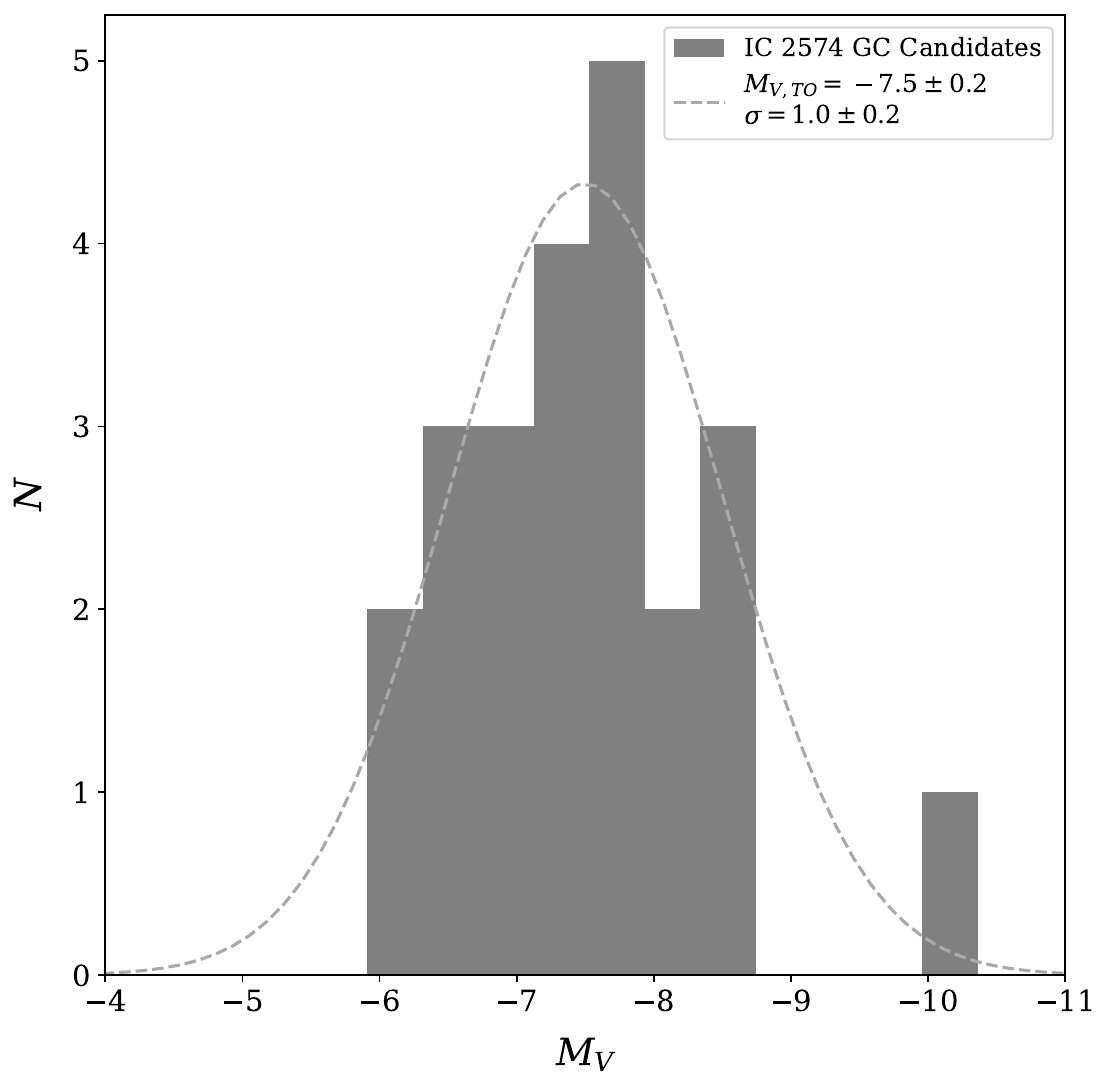}
    \caption{The luminosity function of the GC candidates for IC~2574, spanning a range $-5.9 \geq M_V \geq -10.4$, with a turnover magnitude of $M_V = -7.5 \pm 0.2$ and standard deviation $\sigma = 1.0 \pm 0.2$; scaled Gaussian fit overlaid. The highest luminosity candidate, IC~2574-01, has $M_V = - 10.4$.}
    \label{fig:lumfunc}
\end{figure}

The GC luminosity function is a robust and predictable feature of a galaxy's GC system \citep{Harris_1991}. The GCLF is expected to exhibit a log-normal distribution, typically centred around a turnover (TO) magnitude of $M_{V,\mathrm{TO}}\sim-7.5$, with the GCLF dispersion systematically widening with increasing galaxy luminosity \citep[e.g.][]{Secker_1992,Reed_1994,Whitmore_1995,Harris_1996,Bica_MWGCs_2006,Richtler_2003_GCLF,Jordan_2007,Villegas_2010,Peacock_2010,Harris_2014,Forbes_OpenQs_2018}, with evidence that the TO magnitude can be less luminous ($M_{V,\mathrm{TO}}\sim-7.3$) for low-mass galaxies \citep{Miller_Lotz_2007}.

We present the GCLF for IC~2574 in Fig.~\ref{fig:lumfunc}, which exhibits the expected features of a log-normal distribution centred around $M_{V,\mathrm{TO}}$. To find $M_{V,\mathrm{TO}}$, we employ a Markov Chain Monte Carlo (MCMC) routine with \texttt{emcee} \citep{ForemanMackey_emcee_2013} to fit a Gaussian distribution to the GCLF. The MCMC routine is a robust approach to explore the parameter space and estimate the optimal values for the mean ($M_{V,\mathrm{TO}}$) and standard deviation ($\sigma_{\mathrm{GCLF}}$) of the raw, unbinned data.
We find $M_{V,\mathrm{TO}} = -7.5 \pm 0.2$ and $\sigma_{\mathrm{GCLF}} = 1.0 \pm 0.2$, not dissimilar to the mean values from the sample of dIrr galaxies in \citet[][$M_\mathrm{{V,TO}} = -7.56 \pm 0.02$, $\sigma_\mathrm{{GCLF}} = 1.23 \pm 0.03$]{Georgiev_2009}. A scaled Gaussian distribution with our $M_{V,\mathrm{TO}}$ and $\sigma$ MCMC values is plotted over the GCLF in Fig.~\ref{fig:lumfunc}.

A distinct feature in IC~2574's GCLF is the highest luminosity GC candidate, IC~2574-01. With $M_V=-10.4$, it is comparable in luminosity to $\omega$ Cen \citep[$M_V=-10.3$,][]{Harris_2010}.
The presence of massive clusters is noteworthy in dwarf galaxies, such as young and similarly massive clusters in nearby starburst dwarfs like NGC~1569 \citep{Ho_NGC1569_1996}, NGC~5253 \citep{deGrijs_NGC5253_2013}, NGC~4449, NGC~4214 \citep{Larsen_NGC4214_NGC4449_2004}, II~Zw~40 \citep{Beck_IIZw40_2002} and WLM \citep{Hodge_WLM_1999}, with some showing signs of mergers.
Since IC~2574 has no evidence of recent mergers \citep{Pellerin_2012,Sorgho_2019,Mondal_2021}, the total disruption of a dwarf galaxy massive enough to possess a nuclear star cluster would be surprising, although we do see nuclear clusters in very low-mass dwarfs \citep{vanDokkum_LumGCs_2018,vanDokkum_LumGCs_2019}.
Massive clusters can also be formed in dense regions of galaxy clusters \citep{vandenBergh_2001}, however given the relatively isolated nature of IC~2574, the alternative is that an intense star formation event \citep{Weisz_2008} could be responsible for the creation of this massive $\omega$ Cen-like GC in its gas-rich environment. Spectroscopic follow-up of the GCs would allow us to determine the chemistry and age of each cluster, providing stringent constraints on GC formation models \citep{Forbes_OpenQs_2018}, and allow us to constrain the nature of IC~2574-01. The stable conditions within the relatively isolated IC~2574 may be key to understanding the formation and survival of such massive clusters.

At the opposite end, we see that the GCLF does not extend all the way to the typically lowest luminosity values for GC systems \citep{Pan_M81GCs_2022,Bonatto_FaintGCs_2008,Crnojevic_EriII_2016,Caldwell_AndI_2017}. This is likely due to either photometric incompleteness for low-luminosity, diffuse clusters, or GC destruction.
Completeness corrections in magnitude for the low-luminosity end of the GCLF were performed in Sec.~\ref{subsec:completeness}, and increase the number of GCs to the corrected $N_{\mathrm{GC}}$ count by 3.
An alternative is that the low-luminosity end of the GCLF can be influenced by the destruction of GCs within the galaxy by
internal dynamical processes over long timescales \citep[see:][]{Barmby_2001_GCLF,Fall_Zhang_2001,Carlson_2001}. These dynamical effects are strongest towards the centre of the galaxy.
High density gas present during intense star formation episodes can also destroy GCs \citep{Kruijssen_2012}, which is relevant for IC~2574 as it recently experienced an intense burst of star formation $500~\mathrm{Myr}-1~\mathrm{Gyr}$ ago \citep{Weisz_2008,Pasquali_2008}. The enhanced tidal interactions from this dense gas-rich environment could contribute to the disruption and dissolution of lower luminosity GCs, as opposed to higher luminosity GCs that tend to be higher mass and dense, making them less susceptible to disruption from their environment \citep{Richtler_2003_GCLF}.
A third possible alternative is that IC~2574 only contains a total of 27 GCs and sampling effects have led to the absence of some lower luminosity GCs.

\subsection{The Specific Frequency of GCs in IC~2574}
\label{subsubsec:specificfrequency}

\citet{Harris_1981} defined the GC specific frequency ($S_N$) of a galaxy as the GC formation efficiency, $N_{\mathrm{GC}}$ per unit galaxy luminosity, allowing us to compare $N_{\mathrm{GC}}$ of different galaxies by normalising it to a galaxy with luminosity $M_V$~=~$-15$. It is defined as 

\begin{equation}
    S_N = N_{\mathrm{GC}} \times 10^{0.4(M_V+15)} \hspace{0.025cm},
\end{equation}
\vspace{0.05cm}

\noindent where $M_V$ is the absolute $V-$band magnitude of the host galaxy. IC~2574 has an apparent V-band magnitude of $V~=~10.87\pm0.03$ \citep{Wenger_SIMBAD_2000,Cook_2014}, giving $M_V$~=~$-17.15 \pm 0.04$. We use $N_{\mathrm{GC}}=27\pm5$ to calculate a value for its GC specific frequency of $S_N$~=~4.0~$\pm$~0.8; the error is propagated from $M_V$ and the Poissonian uncertainty in $N_{\mathrm{GC}}$. 

\begin{figure*}
    \centering
        \includegraphics[width=\columnwidth]{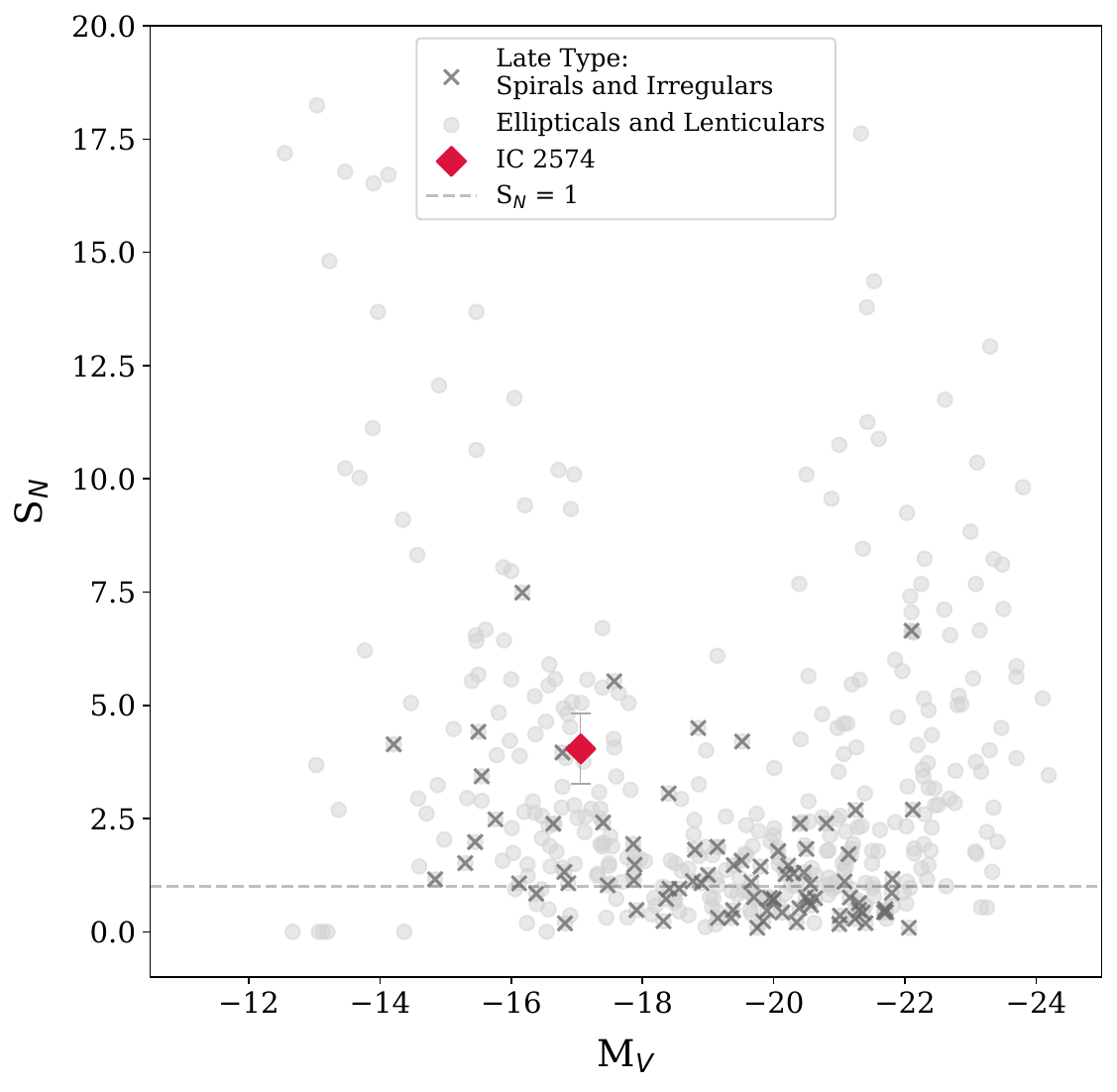}
    \includegraphics[width=\columnwidth]{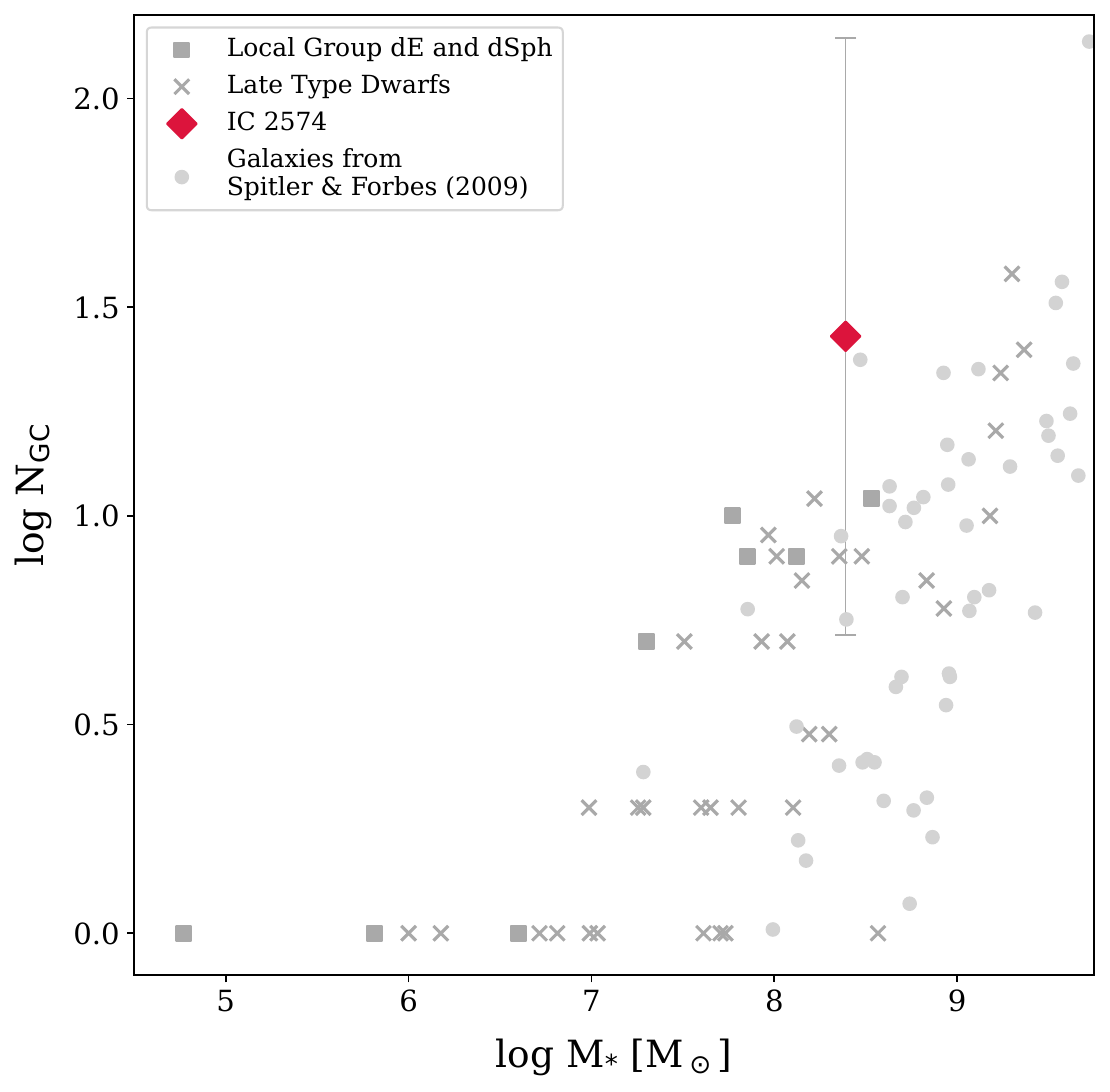}
    \caption{\textit{Left}: The specific frequency $S_N$ vs. absolute $V-$band magnitude $M_V$ of galaxies. Cross symbols represents late-type galaxies (spirals and irregulars), and round symbols represent early-type galaxies (ellipticals and lenticulars). IC~2574 is plotted as a red diamond. Comparing IC~2574 to other late-type dwarfs (cross symbols) shows us that it is consistent with the trend of increased $S_N$ with decreasing luminosity for dIrrs, and that it exists on the higher end of this upward turn of the distribution. \textit{Right}: The related distribution of $N_{\mathrm{GC}}$ vs. $M_*$. Square symbols are Local Group dE and dSph and the cross symbols are late-type dwarfs from \citet{Forbes_2018_Extending}. Round symbols are galaxies from \citet{Spitler_2009}, dominated by early-type massive ellipticals and lenticulars. Dwarfs for which $N_{\mathrm{GC}}=0$ are not included. IC~2574 is plotted as a red diamond. IC~2574 has a higher $N_{\mathrm{GC}}$ (and thus $S_N$) on this relation compared to other late-type dwarfs (cross symbols) within its luminosity band, hinting comparatively more efficient GC formation because of its unique conditions.}
    \label{fig:SN_NGC}
\end{figure*}

Galaxy morphology, environment, SFH and luminosity are all factors to consider for $S_N$ \citep{Harris_1981,Peng_2008, Harris_2013,Lim_2018}. Fig.~\ref{fig:SN_NGC} shows the relation between $S_N$ and $M_V$ with the galaxy catalogue from \citet{Harris_2013} and IC~2574 overlaid. The key feature of this relation is the distinct U-shaped distribution, discussed extensively in \citet{Forbes_2005}, \citet{Peng_2008} and \citet{Georgiev_2010}, where galaxies with intermediate luminosities gather tightly in the `valley' of the distribution which occurs around $S_N\simeq$ 1, with the dwarf and giant galaxies at either end scattering greatly upwards with much higher average $S_N$; this characteristic shape is driven by the halo mass of the galaxies and essentially the shape of the stellar mass-halo mass relation \citep[SHMR;][]{Wechsler_2018,Girelli_2020}, as there is a near-constant $M_{\mathrm{GC}}-M_{200}$ relation and uniformity in the GCLF of the galaxies across the sample. 

From Fig.~\ref{fig:SN_NGC}, IC~2574 appears to have a typical $S_N$ for its luminosity (albeit at the higher end of the range for late-type galaxies), consistent with a clear trend of other late-type dIrrs having increased $S_N$ with decreasing luminosity \citep{Georgiev_2010}. On the right of Fig.~\ref{fig:SN_NGC} we plot the related distribution of $N_{\mathrm{GC}}$ against the stellar mass, where the stellar mass for IC~2574 is $\log M_* = 8.39 \hspace{0.05cm} M_{\odot}$ \citep{Lee_2006}. This shows that the $N_{\mathrm{GC}}$ of IC~2574 is on the higher end of the distribution than that of other late-type dwarfs within the same luminosity. 
High specific frequencies of late-type dwarfs can be attributed to bursty star formation, consistent with the recent burst seen in IC~2574 $\sim$ 1 Gyr ago \citep{Pasquali_2008,Weisz_2008}.
The $S_N$ vs. $M_V$ relation in the low-mass regime indicates that dwarf galaxies tend to have a greater fraction of their stellar mass in GCs as their luminosity decreases. One possible explanation for this trend is that dwarf galaxies with GCs are more efficient at forming and retaining these clusters, or alternatively, they are less efficient at forming field stars.
IC~2574's position on this relation demonstrates excellent agreement with the expected behaviour for a galaxy of its specific morphology and environment.

\subsection{GC System-Galaxy Scaling Relations}
\label{subsec:scaling}

The relationship between a GC system and its host galaxy is complex and multifaceted, requiring the study of many different aspects of both the GC system and the host galaxy itself. Here we calculate $M_{\mathrm{GC}}$ of IC~2574 by multiplying $N_{\mathrm{GC}}$ with an average GC mass of $4\times10^5 M_{\odot}$, as in \citet{Spitler_2009}. This gives us log~$M_{\mathrm{GC}} = 7.03 \pm 0.08$. We plot IC~2574 on the GC system mass-stellar mass relation and the GC system mass-halo mass relation. These are recreated from \citet{Forbes_2018_Extending} with their Tables 1 and 2, which are Local Group dE and dSph galaxies and late-type dwarfs respectively. We also use the catalogue of galaxies from \citet{Spitler_2009}, which is dominated by early-type massive elliptical and lenticular galaxies. 

\subsubsection{GC System Mass-Stellar Mass Relation}
\label{subsubsec:GCSM-stellar}

\begin{figure}
    \centering
    \includegraphics[width=\columnwidth]{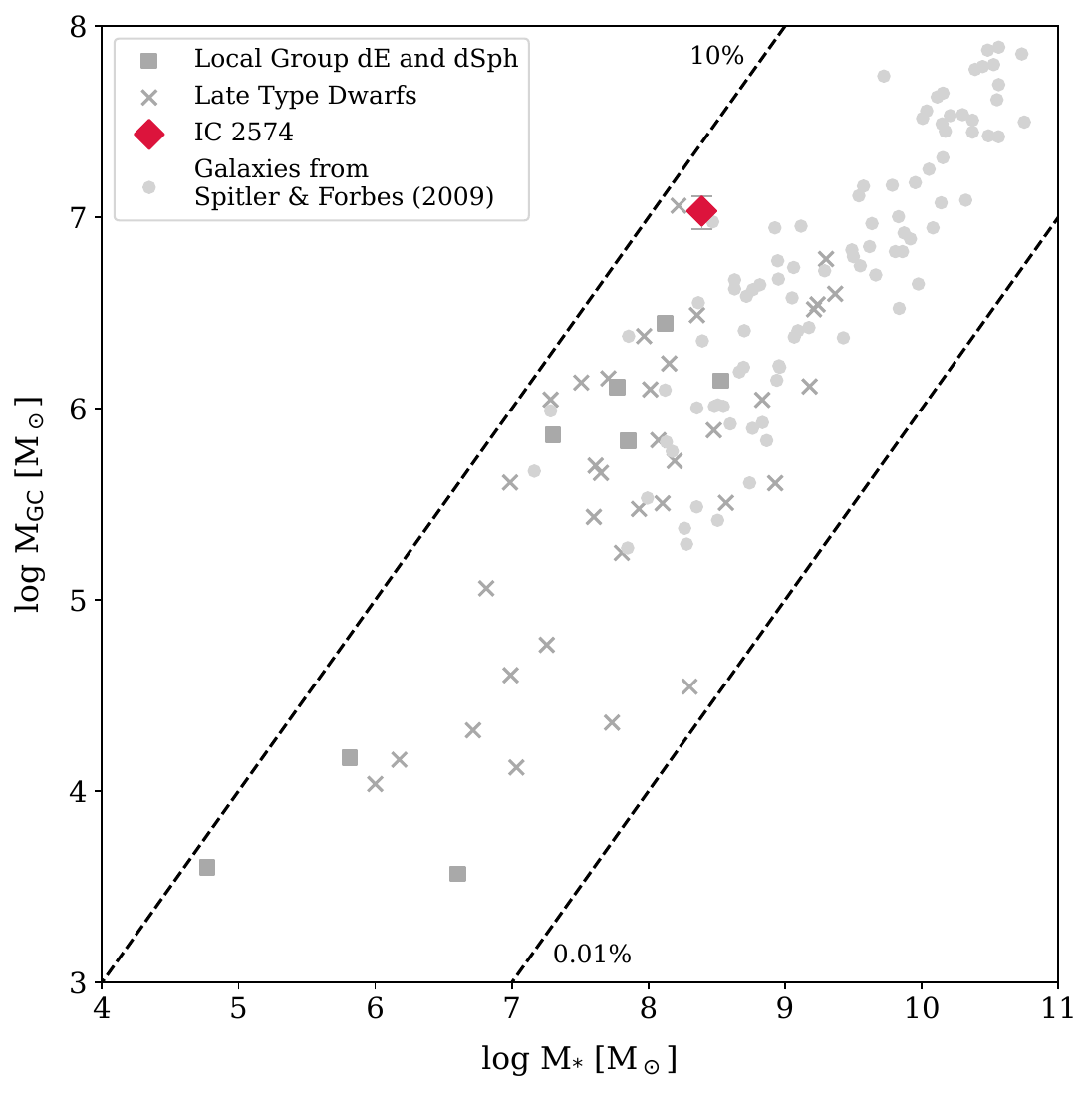}
    \caption{$M_{\mathrm{GC}}$ as a function of stellar mass $M_*$ of each galaxy. IC~2574 is plotted as a red diamond. Square symbols are Local Group dE and dSph and cross symbols are late-type dwarfs from \citet{Forbes_2018_Extending}, round symbols are galaxies from the \citet{Spitler_2009} catalogue, dominated by early-type massive ellipticals and lenticulars. Dashed lines represent 10\% and 0.01\% ratios of $M_{\mathrm{GC}}$ to $M_*$. IC~2574 sits high on this relation, close to the 10\% boundary, suggesting efficient GC formation in this galaxy.}
    \label{fig:gcsmsm}
\end{figure}

We begin with the physically motivated GC system mass-stellar mass relation, shown in Fig. \ref{fig:gcsmsm}.
We find that IC~2574 is consistent with what is expected from its stellar mass, lying sensibly within the boundaries in the dashed lines of Fig.~\ref{fig:gcsmsm}. The boundaries represent GC system mass to galaxy stellar mass ratios of 10\% and 0.01\%; IC~2574 sits close to the 10\% boundary, suggesting that this galaxy has efficient GC formation relative to its (field) star formation.

\subsubsection{GC System Mass-Halo Mass Relation}
\label{subsubsec:GCSM-halo}

\begin{figure*}
    \centering
    \includegraphics[width=\textwidth]{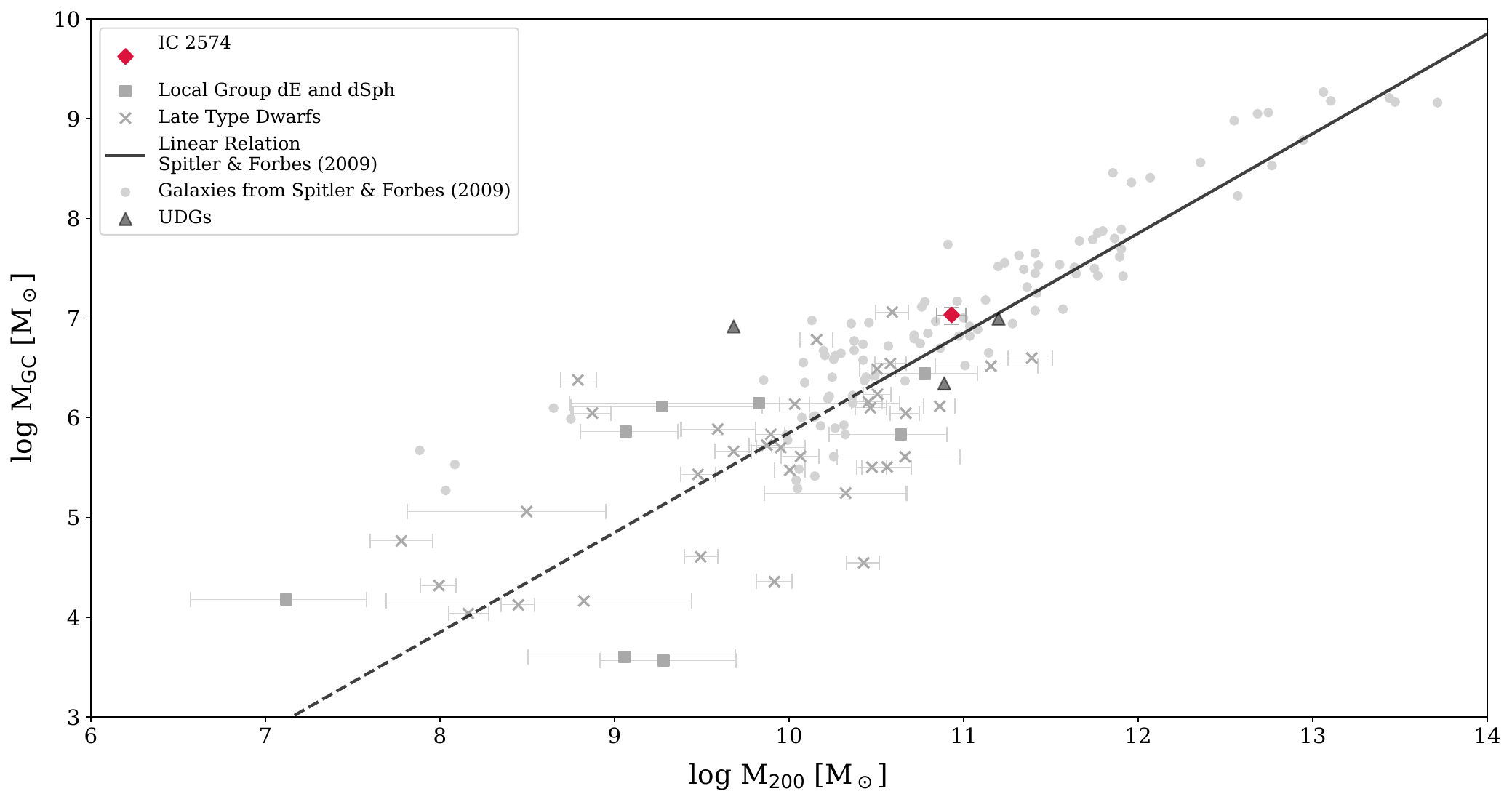}
    \caption{The GC system mass-halo mass relation, $M_{\mathrm{GC}}-M_{200}$. Square symbols are Local Group dE and dSph, cross symbols are late-type dwarfs from \citet{Forbes_2018_Extending}, round symbols are from a sample of mostly early-type massive elliptical and lenticular galaxies from \citet{Spitler_2009}. Three UDGs have been plotted as in \citet{Forbes_2018_Extending}: VCC 1287 \citep{Beasley_2016}, DF 44 \citep{vanDokkum_DF44_2019} and DF 2 \citep{vanDokkum_DMD_2018}. The red diamond is IC~2574, which falls just on the $M_{\mathrm{GC}}-M_{200}$ relation. The solid line is the linear fit from \citet{Spitler_2009}. IC~2574 has a GC system mass expected of a galaxy with its halo mass, from independent measures of both its halo mass, calculated from its kinematic data and its GC system.}
    \label{fig:gcsmhm}
\end{figure*}

In Fig. \ref{fig:gcsmhm} we show the GC system mass-halo mass relation with IC~2574 overlaid. The data for other galaxies is assembled from \citet{Spitler_2009} and \citet{Forbes_2018_Extending}. The halo mass of the host galaxies from \citet{Forbes_2018_Extending} have been calculated from their kinematics. We also plot the three UDGs as in \citet{Forbes_2018_Extending}; VCC 1287 \citep{Beasley_2016}, DF 44 \citep{vanDokkum_DF44_2019} and DF 2 \citep{vanDokkum_DMD_2018}.
These three UDGs have stellar masses of the order $M_* \sim 10^7 - 10^8 M_\odot$, similar to IC~2574. This spread in their position on the GC system mass-halo mass relation is interesting to note as a comparison to IC~2574, which lies on the boundary of UDG characteristics.

In order to add IC~2574 to the GC system mass-halo mass relation, we make an independent empirically derived estimate of $M_{200}$ using its $v_{\mathrm{max}}$ of $77.6 \pm 5.0 \hspace{0.05cm} \mathrm{kms}^{-1}$ \citep{Oh_2011}. 
We use the relations (from e.g. \citealt{Sigad_2000});

\begin{equation}
    M_{200} = \frac{4\pi}{3} {R_{200}}^3 \times 200 \hspace{0.1cm} \rho_{\mathrm{cm}},
\end{equation}

\begin{equation}
    V_{200} = \sqrt{\frac{GM_{200}}{R_{200}}},
\end{equation}

\noindent and the Navarro–Frenk–White (NFW) profile for the density distribution \citep{NFW_1996}:

\begin{equation}
    M_{200} = 4\pi\rho_0 {R_s}^3 \Big[ 
\ln(1+c) - \frac{c}{1 + c} \Big] ,
\end{equation}

\noindent with the concentration parameter $c = R_{200}/R_s$ \citep{Diemer_2019}, values for $c$ taken from \citet{Castellanos_2020}, and the relation $R_{\mathrm{max}} = 2.16 \times R_s$ \citep{Weinmann_2013}, to rearrange for known parameters and substitute in to the below equation (e.g. \citealt{Dutton_2014,Comparat_2017}), 

\begin{equation}
    v_{\mathrm{max}} = \sqrt{\frac{GM(<R_{\mathrm{max}})}{R_{\mathrm{max}}}}
\end{equation}

\noindent where we rearrange $v_{\mathrm{max}}$ for $M_{200}$. We estimate IC~2574's halo mass as $M_{200} = (8.54 \pm 1.65) \times 10^{10}$ $M_{\odot}$ (log $M_{200}$ = $10.93\pm0.08$ $M_{\odot}$), with the uncertainties propagated from $v_{\mathrm{max}}$, where we assume the uncertainty stated in Table~\ref{tab:ic2574properties} is the velocity difference between the true and derived rotation velocities, which is the maximum uncertainty from limitations in the instrument.

We also calculate the value for $M_{200}$ using the SHMR abundance matching technique from \citet{Read_2017}, with $M_*$ from Table~\ref{tab:ic2574properties}, finding $M_{200} = 5.2 \pm 0.4 \times 10^{10} M_\odot$ (log $M_{200}$ = $10.72\pm0.03$ $M_{\odot}$). While the empirical kinematic measurement and the abundance matching SHMR prediction yield $M_{200}$ estimates that do not have overlapping uncertainty ranges, both values are well within the same order of magnitude.

\citet{Burkert_Forbes_2020} describe the relationship\footnote{This is used as an approximation since $M_{\mathrm{vir}}$ is not necessarily equal to $M_{200}$; it represents the mass enclosed within the virial radius, at which the density is equal to the critical density $\rho_{c}$ of the Universe, multiplied by an overdensity constant $\Delta_{c}$. If $\Delta_{c}$ = 200, which is the typical assumption, then it becomes $M_{200}$.} between $N_{\mathrm{GC}}$ and $M_{\mathrm{vir}}$ as log $N_{\mathrm{GC}}=-9.58\hspace{0.05cm}(\pm 1.58)+0.99\hspace{0.05cm}(\pm 0.13)~\times~\mathrm{log}M_{\mathrm{vir}}(M_{\odot})$, where our $N_{\mathrm{GC}}=27\pm5$ gives a value for log $M_{\mathrm{vir}} = 11.1 \hspace{0.05cm}\mathrm{M}_{\odot}$, which has an uncertainty of 0.3
dex \citep{Burkert_Forbes_2020}. This shows agreement with the empirically derived halo mass above.\\

We have used the empirically derived $M_{200}$ to add IC~2574 to the $M_{\mathrm{GC}}-M_{200}$ relation in Fig. \ref{fig:gcsmhm}. We see that IC~2574 agrees with the near-linear observed relation from \citet{Spitler_2009}, well within its scatter, which is what we would expect given the luminosity of this galaxy. If confirmed, IC~2574 does appear to have a substantial GC system. This result also agrees with simulated GC formation scenarios such as those of \citet[][and their previous works cited within]{Chen_2023}, who account for the stochasticity of GC occupation in the low-mass regime, which arises from a complex interplay of baryonic effects. 
They predict the near-linearity of the $M_{\mathrm{GC}}-M_{200}$ relation down to $M_{200} \simeq 10^8 M_\odot$. 
\citet{Chen_2023} also predict that more than 50\% of galaxies with $M_* \leq 2 \times 10^7 M_\odot$ do not host GCs. \citet{Eadie_2022} use
a statistical model to find the 50\% probability point for a galaxy to host GCs is $M_* \leq 10^{6.8} M_{\odot}$; they discuss improvements to this value by incorporating a hierarchical model to account for uncertainties in $M_*$ \citep{Berek_2023}, and ultimately the observational need to add galaxy data in and below the transition region between the observed relation ($M_{\mathrm{GC}}-M_*$ instead, due to limitations in independent $M_{200}$ measurements) and where galaxies appear to host no GCs.
This motivates the critical next steps in building the census of galaxies in the low-mass regime with secure halo mass measurements, especially that of relatively isolated dwarfs like IC~2574 to have the effects of tidal interactions minimised on their halos and GC systems, in order to understand their GC formation efficiencies in a fundamental way. We can use similar techniques to add lower mass galaxies to this relation and test GC formation models that have made predictions to lower masses \citep{Muratov_Gnedin_2010,Li_Gnedin_2014,Choksi_2018,Chen_2022,Eadie_2022,Chen_2023}. Increasing this low-mass census will assess the robustness of using the GC system mass-halo mass relation as a technique for halo mass estimates, as this near-linear correlation could be evidence of universal physical processes governing GC formation and evolution in galaxies across all mass scales.

\section{Conclusions}
\label{sec:conclusion}

In this paper we have presented a quantitative description of the GC population of IC~2574 for the first time. We summarise our key findings below.

\begin{itemize}
    \item We have surveyed deep HST/ACS images for IC~2574 and identified 23 candidate globular clusters for this system. After completeness corrections, we find that IC~2574 has a total GC system of $N_{\mathrm{GC}}~=~27~\pm~5$. Their luminosities range from $-5.9 \geq M_V \geq -10.4$, and their sizes are typical of GCs, spanning a range of $1.4 \leq r_h \leq 11.5$~pc.

    \item Our 23 candidate GCs show evidence of colour bimodality. This could suggest two distinct formation channels for these GCs. Our 5 excluded candidates exhibit extremely blue colours, which could potentially be young massive star clusters linked to the burst of intense star formation in IC~2574 $\sim$1~Gyr ago.

    \item We find a very high-luminosity GC candidate in IC~2574 (IC~2574-01) that is an analogue of the unusual and rare $\omega$ Cen GC in the Milky Way. If confirmed spectroscopically, this makes IC~2574 a notable addition to a collection of dwarfs that host high-luminosity massive GCs, and can help us understand whether it has had a significant past merger with a galaxy containing a nuclear star cluster. 
    
    \item Although IC~2574 is not technically a UDG it is a fairly unusual galaxy in that it has a large effective radius and low surface brightness. This could be an interesting note in the discussion of UDGs and their high-luminosity GCs if we can confirm the membership and nature of IC~2574-01.

    \item We find that IC~2574's GC system is in agreement with that expected for a galaxy of its mass and morphology. It has a GC specific frequency of $S_N=4.0\pm0.8$, consistent with other late-type dIrrs. It sits on the GC system mass-stellar mass distribution ($M_{\mathrm{GC}}-M_*$) with a high ratio of $\sim$10\%, suggesting efficient GC formation relative to its (field) star formation. Its position in the $N_{\mathrm{GC}}$ and $M_*$ distribution is a related illustration that demonstrates IC~2574 hosts more GCs than is typical for a galaxy of its stellar mass.
    
    \item IC~2574 experienced a recent intense burst of star formation $\sim$1 Gyr ago that accounts for 15\% of the stars formed, and which might explain the 5  extremely blue star clusters we found. 
    These 5 objects make up 16\% of the total number of candidate clusters. 

    \item IC~2574 has a well-determined HI rotation curve, which we use to derive its halo mass and place it on the GC system mass-halo mass relation ($M_{\mathrm{GC}}-M_{200}$). IC~2574 agrees very well with this relation, suggesting it has a fairly normal GC population in terms of the GC system mass, although it does host the unusually high-luminosity $\omega$ Cen analogue IC~2574-01.
\end{itemize}

For future work it is paramount that we confirm this population of candidate GCs with spectroscopic data to obtain their radial velocities, metallicities, estimate their ages and obtain their chemical abundance measurements. This will enable us to truly understand why the distribution of colours and sizes within the GC system are this way and build a broader picture of what happened in IC~2574 from the formation of its ancient population of GCs to its possible young star cluster formation in its recent star formation episode.

\section*{Acknowledgements}
We thank the referee for their useful and detailed comments and suggestions that helped to improve the quality of this work.
This work was carried out based on observations made with the NASA/ESA Hubble Space Telescope, and obtained from the Hubble Legacy Archive, which is a collaboration between the Space Telescope Science Institute (STScI/NASA), the Space Telescope European Coordinating Facility (STECF/ESA) and the Canadian Astronomy Data Centre (CADC/NRC/CSA). The authors made use of Astropy, a community-developed core Python package for
Astronomy \citep{Astropy_2013}, the data visualisation library Matplotlib \citep{Hunter_Matplotlib_2007}, NumPy \citep{Harris_NumPy_2020}, SciPy \citep{Virtanen_SciPy_2020}, pandas \citep{McKinney_pandas_2010}, the MCMC sampler \texttt{emcee} \citep{ForemanMackey_emcee_2013}, the astronomical imaging and data visualisation program SAOImage DS9 \citep{DS9_2000,Joye_DS9_2003} the NASA/IPAC Extragalactic Database (NED, which is operated by the Jet Propulsion Laboratory, California Institute of Technology, under contract with the National Aeronautics and Space Administration) and the SIMBAD database, operated at CDS, Strasbourg, France \citep{Wenger_SIMBAD_2000}. The authors also made extensive use of \texttt{Source Extractor} and subsequent user manuals \citep{Bertin_Arnouts_1996,Holwerda_2005}. 

\section*{Data Availability}

The data underlying this article will be shared on reasonable request to the corresponding author.




\bibliographystyle{mnras}
\bibliography{ref}



\bsp	
\label{lastpage}
\end{document}